\documentclass[pdflatex,sn-nature]{sn-jnl}

\usepackage{graphicx}%
\usepackage{multirow}%
\usepackage{amsmath,amssymb,amsfonts}%
\usepackage{amsthm}%
\usepackage{mathrsfs}%
\usepackage[title]{appendix}%
\usepackage{xcolor}%
\usepackage{textcomp}%
\usepackage{manyfoot}%
\usepackage{booktabs}%
\usepackage{bm}%
\usepackage{algorithm}%
\usepackage{algorithmicx}%
\usepackage{algpseudocode}%
\usepackage{listings}%
\usepackage{hyperref}
\usepackage{xr-hyper}

\theoremstyle{thmstyleone}%
\theoremstyle{thmstyletwo}%

\theoremstyle{thmstylethree}%

\unnumbered

\newcommand{\eqn}[1]{eqn.~(\ref{#1})}

\newcommand{\fig}[1]{Figure~\ref{#1}}

\newcommand{\tab}[1]{Table~\ref{#1}}

\newcommand{\msec}[1]{Methods section ``\nameref{#1}''}
\newcommand{\msecn}[1]{``\nameref{#1}''}
\theoremstyle{definition}
\newcommand{\angstrom}{\mbox{\normalfont\AA}}
\newcommand{\rev}[1]{{{#1}}}
\renewcommand{\figurename}{\rev{Figure}}
\hypersetup{
    colorlinks=true,
    hypertexnames=false,
    linkcolor=blue,
    filecolor=blue,
    urlcolor=blue,
    citecolor=blue,
    pdftitle={main},
    pdfauthor={Moon-ki Choi},
}

\begin{document}

\title[Unified Learning of Interatomic Potentials and Electronic Hamiltonians, with Applications in 2D Materials]{Unified Learning of Interatomic Potentials and Electronic Hamiltonians, with Applications in 2D Materials}


\author[1,2]{\fnm{Moon-ki} \sur{Choi}}\email{choi0652@illinois.edu}

\author[3]{\fnm{Daniel} \sur{Palmer}}\email{dpalmer3@illinois.edu}

\author[1,2]{\fnm{Shivam} \sur{Sharma}}\email{shivams@illinois.edu}

\author*[1,2]{\fnm{Harley~T.} \sur{Johnson}}\email{htj@illinois.edu}

\affil*[1]{\orgdiv{Department of Mechanical Science and Engineering}, \orgname{University of Illinois Urbana--Champaign}, \orgaddress{\street{1206 W. Green St.}, \city{Urbana}, \postcode{61801}, \state{Illinois}, \country{USA}}}

\affil[2]{\orgdiv{Materials Research Laboratory}, \orgname{University of Illinois Urbana--Champaign}, \orgaddress{\street{104 South Goodwin Ave.
MC-230}, \city{Urbana}, \postcode{61801}, \state{Illinois}, \country{USA}}}

\affil[3]{\orgdiv{Department of Materials Science \& Engineering}, \orgname{University of Illinois Urbana--Champaign}, \orgaddress{\street{1304 W Green St}, \city{Urbana}, \postcode{61801}, \state{Illinois}, \country{USA}}}


\abstract{We introduce UnIPH, an equivariant graph neural network designed to jointly predict both interatomic potentials and tight-binding (TB) Hamiltonians for an atomic structure.
UnIPH is trained using density functional theory (DFT) calculations followed by Wannier projection to predict energies and forces as node-level targets and Wannier-projected TB Hamiltonians as edge-level targets. This enables physically consistent modeling of coupled mechanical–electronic responses with near-DFT-level accuracy. \rev{After separately training two UnIPH models on DFT data for bilayer graphene and monolayer MoS$_2$, respectively, we find that these models capture} key deformation–electronic effects: in twisted bilayer graphene, \rev{the trained UnIPH} reveals how interlayer spacing, in-plane strain, and out-of-plane corrugation drive isolated flat-band formation, and further shows that modulating substrate interaction strength can generate flat bands even away from the magic angle ($\sim 1.1^\circ$). For monolayer MoS$_2$, UnIPH accurately reproduces phonon dispersions and strain-dependent band-gap evolution, and further predicts how a non-uniform substrate creates localized strain pockets that spatially modulate the conduction- and valence-band edges.}




\maketitle

\section{Introduction}\label{sec:intro}
Two-dimensional (2D) materials can withstand significant mechanical deformation without undergoing structural failure. This property enables strain engineering, a powerful technique in which mechanical strain is used to modulate the electronic, optical, magnetic, transport, and mechanical properties of 2D materials \cite{akinwande2017,peng2020,pandey2023,katiyar2025}. A well-known example is MoS$_2$, a transition metal dichalcogenide (TMD) material in which uniform strain and bending can be used to tune the band gap \cite{conley2013,petHo2019,tang2022,xiao2014,na2022}. Diverse strategies have been developed to induce strain in 2D materials, using non-flat substrates \cite{zhang2021,shin2024,li2020}, nano-indentation \cite{jiang2020}, and stressors placed on top of the 2D materials \cite{zhang2024}. When the strain field is spatially non-uniform, these approaches can create localized band-edge landscapes in TMD monolayers, offering a route to spatially patterned electronic and optical functionality \cite{castellanos2013,hui2014,li2015}.

In addition to the strain engineering, layer stacking and twisting offer another route to modulate the electronic structure of 2D materials. A prominent example is twisted bilayer graphene (TBG), where rotating two graphene layers relative to each other by an angle near the “magic angle” ($\sim1.1^\circ$) leads to isolated flat bands in the band structure. This leads to various strongly correlated phases such as Mott insulating, ferromagnetic, and superconducting \cite{cao2018,lu2019,andrei2020}. In TBG, structural relaxation-induced deformation also plays a crucial role in shaping the electronic structure \cite{guinea2019,kazmierczak2021,carr2018,krongchon2023}. Such deformation includes in-plane strain \cite{letb,sung2022}, out-of-plane corrugation \cite{rakib2022,krongchon2023}, and variations in interlayer distance \cite{carr2018}, all of which contribute to the resulting electronic landscape. 

Modeling these deformation-induced electronic states requires coupling structural mechanics with electronic-state calculations. Although density functional theory (DFT) is a powerful tool for studying such deformation-driven phenomena, its high computational cost limits routine application to relatively small systems \cite{bowler2012methods}. For larger atomistic structures, such as magic-angle TBG, an alternative is a sequential workflow in which an interatomic potential (IP) relaxes the structure and a tight-binding (TB) model is then used to compute the electronic structure \cite{rakib2022,trambly2010,letb,moon2012}. However, because the IP and TB models are typically developed independently, this approach can lack consistency between the mechanical and electronic descriptions and may not reliably capture their coupled response. It is also difficult to adopt such an approach in new or poorly studied materials, where matched IP and TB models are often unavailable.

\rev{For large atomistic structures, density-functional tight binding (DFTB) is a suitable approach for investigating energies, forces, and band structures at lower computational cost than DFT \cite{elstner1998self}. Machine-learning extensions of tight-binding models are also emerging, for example through the optimization of tight-binding parameters or the adaptation of Slater--Koster parameters to local atomic environments \cite{mcsloy2023tbmalt,song2026adaptive}. In conventional DFTB, however, the hopping parameters are represented by Slater--Koster functions that depend only on interatomic distance, so changes in the many-body atomic environment are not captured explicitly.} 

Recent advances in machine learning models, particularly in equivariant graph neural networks (GNNs), offer promising solutions for predicting a broad range of properties in atomistic systems. A prominent example is machine-learning interatomic potentials (MLIPs) such as NequIP \cite{nequip}, MACE \cite{mace}, GAP \cite{bartok2010}, and Allegro \cite{musaelian2023}, which predict atomic forces and energies at DFT-level accuracy. \rev{GRACE extends the atomic cluster expansion to a graph basis and provides a systematic connection to equivariant message-passing interatomic potentials \cite{bochkarev2024graph,lysogorskiy2026graph}.} Other models, such as CHGNet \cite{deng2023}, extend this capability to include magnetic degrees of freedom by incorporating charge and spin information. Beyond these IPs, GNN-based approaches have also been developed to learn the Kohn–Sham Hamiltonian from linear-combination-of-atomic-orbitals (LCAO) DFT calculations \cite{deeph,deephe3}, enabling accurate band-structure predictions for an atomic structure. 

\rev{Other machine-learning approaches instead fit environment-dependent Slater--Koster bond integrals \cite{wang2025gputb} or tight-binding parameters for specific defect structures \cite{schattauer2022machine}.}
Although these approaches provide either mechanical or electronic information, predicting deformation-induced electronic structures in large-scale structures with near-DFT accuracy requires a machine-learning framework capable of learning coupled mechanical and electronic responses across diverse atomic configurations.

\rev{In this study, UnIPH builds on e3nn-based equivariant GNN frameworks and jointly learns and predicts an interatomic potential and a Wannier tight-binding Hamiltonian from shared representations \cite{e3nn}.} \rev{Its distinctive task formulation is a unified multiphysics model in which a single $E(3)$-equivariant GNN jointly learns two targets from the same DFT reference workflow for each atomic configuration: a potential-energy surface from total energies and forces, and an environment-dependent TB Hamiltonian reconstructed from the corresponding DFT electronic structure through Wannier90. This common reference enables mechanically and electronically consistent, DFT-level predictions within one model.} This model is trained on DFT reference data; total energies and atomic forces are used as training targets for IP prediction, whereas Wannier-projected TB matrices provide the targets for predicting TB Hamiltonian \cite{wannier90}. In the graph representation, node features are used to predict atomic energies and forces, whereas edge features are used to predict TB Hamiltonian. This unified multitask learning scheme not only enables a single GNN architecture to learn and predict both mechanical and electronic properties in a physically consistent manner, but also improves efficiency by leveraging shared representations, thereby maintaining near-DFT accuracy across the tasks \cite{multitask}.
\rev{The graphene and MoS$_2$ applications presented below use independently trained, material-specific models.}

We first train UnIPH on bilayer graphene systems to investigate structural relaxation and its impact on the electronic band structure of TBG. To validate the model, we compare UnIPH predictions against DFT results for interlayer interaction energies, generalized stacking-fault energies (GSFE), and electronic band structures in AB stacking, AB stacking with a perturbation, and large-angle TBG configurations. UnIPH is then applied to relax large-scale TBG structures and compute their electronic band structures across a range of twist angles and substrate conditions. Key electronic features (bandwidth, electron gap, and hole gap) are analyzed along with structural deformations (interlayer distance, in-plane strain, and out-of-plane corrugation). Finally, we show that the presence and tunable strength of substrate interactions significantly influence structural relaxation and, in turn, electronic properties, enabling the emergence of isolated flat bands even at non-magic twist angles.

We also train UnIPH on monolayer MoS$_2$ using energies, forces, and TB Hamiltonians obtained from DFT calculations and Wannier projections. The model accurately reproduces a phonon dispersion and band structures under uniform strain, predicting a band-gap evolution across various combinations of 2D strain components ($\varepsilon_{xx}$, $\varepsilon_{yy}$, and $\varepsilon_{xy}$). Furthermore, by relaxing a large MoS$_2$ supercell on a non-uniform continuum substrate, we show that substrate-induced local strain creates a spatially varying band-edge landscape with localized band-gap narrowing.

\section{Results}\label{sec:model}
\subsection{Overview of unified training and inference}
\begin{figure}[h!]
\includegraphics[width=0.95\textwidth]{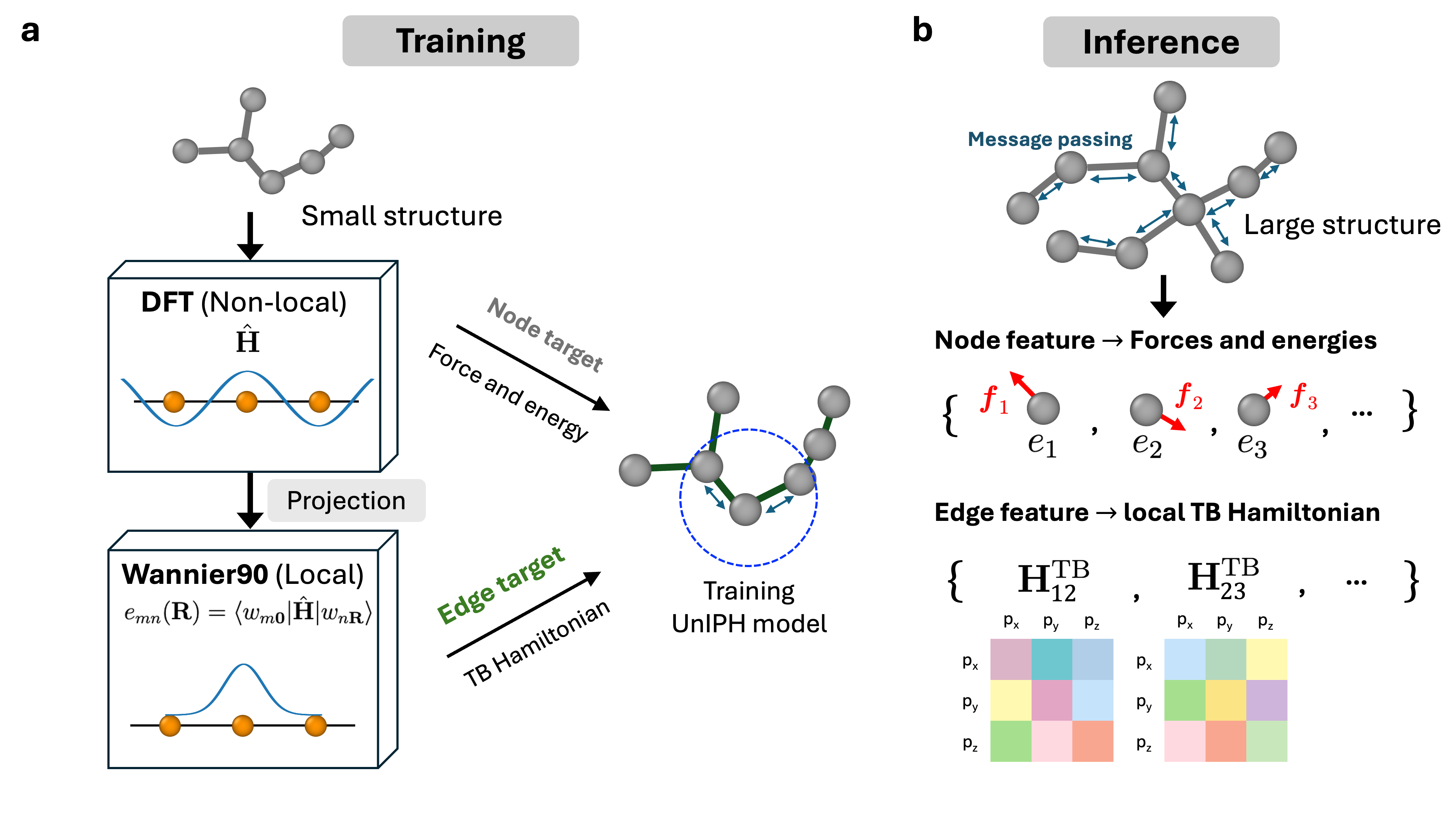}
\caption{
Overview of the training and inference workflow of UnIPH.
(a) Data generation pipeline for training UnIPH. For each atomic structure, DFT calculations provide total energy, atomic forces, and the Kohn-Sham Hamiltonian ($\hat{\mathbf{H}}$). Then $\hat{\mathbf{H}}$ is projected onto a localized Wannier basis to generate TB matrix elements for each pair of orbitals. These energies and forces (node targets), and TB matrices (edge targets) are jointly used to train UnIPH.
(b) Inference stage using the trained UnIPH model. For an atomic structure, the model predicts both the interatomic potential (red arrows for predicted forces) from node features, and the local TB Hamiltonians $\mathbf{H}_{ij}^{\rm TB}$ from edge features. As an example, we present $p$-orbital systems, where each $\mathbf{H}_{ij}^{\rm TB}$ is a $3 \times 3$ matrix. The color in the matrix visualization represents variation of the magnitude and sign for each element. \rev{Molecular structures in this figure were visualized using OVITO version 3.7.6.}}
\label{fig:schematic}
\end{figure}

\fig{fig:schematic} presents an overview of the training and inference workflow in UnIPH. \fig{fig:schematic}a illustrates the construction of the training set using DFT calculations followed by Wannier projections. For each perturbed or strained atomic configuration, a self-consistent field DFT calculation is performed (see \msec{sec:method:dft} for computational details). These calculations result in the total energy and atomic forces, which serve as reference values for node-level prediction during training. The DFT output also includes the Kohn–Sham Hamiltonian ($\hat{\mathbf{H}}$) expressed in a non-local basis (plane-wave basis), which encodes the electronic structure of the system. 
Directly training on $\hat{\mathbf{H}}$ is infeasible due to the local nature of message passing in the graph neural network \cite{deeph}. To address this, $\hat{\mathbf{H}}$ is projected into a localized representation that captures real space orbital interactions. This is achieved by employing Wannier90 \cite{wannier90} to project the DFT Hamiltonian ($\hat{\mathbf{H}}$) onto a set of maximally localized Wannier functions, thereby enabling a compact and physically transparent TB description. The resulting Wannier-projected TB Hamiltonian elements are defined as
\begin{equation}
e_{mn}(\mathbf{R}) = \langle w_{m\mathbf{0}}\vert \hat{\mathbf{H}}\vert w_{n\mathbf{R}} \rangle,
\label{eqn:dft_w90}
\end{equation}
where $e_{mn}(\mathbf{R})$ denotes a real-space hopping energy from orbital $n$ in the unit cell located at lattice vector $\mathbf{R}$ to orbital $m$ in the reference (origin) unit cell. Here, $w_{m\mathbf{0}}$ and $w_{n\mathbf{R}}$ are Wannier functions associated with orbitals $m$ and $n$ in their respective unit cells. Further details of the Wannier projection process for bilayer graphene and monolayer MoS$_2$ are provided in \msec{sec:method:wannier}.
Using \eqn{eqn:dft_w90}, we construct a TB Hamiltonian that reproduces the DFT electronic structure using a significantly smaller set of localized basis functions. For each pair of atoms $i$ and $j$, a local TB Hamiltonian block $\mathbf{H}_{ij}^{\mathrm{TB}}$ is assembled from the corresponding Wannier-projected matrix elements. These computed local TB Hamiltonian matrices, associated with atom pairs (edges), serve as reference values for edge-level predictions during training. Because both the node targets (atomic forces and energies) and the edge targets (local TB Hamiltonian matrices) are derived from the same DFT calculation, a well-trained UnIPH model should achieve DFT-level accuracy for both the potential energy surface and the electronic structure while maintaining consistency between the predicted mechanical and electronic responses.

\fig{fig:schematic}b illustrates the inference stage of UnIPH. For an atomic structure, potentially much larger than those in the training set, UnIPH predicts atomic forces and energies from node features, and local TB Hamiltonians $\mathbf{H}_{ij}^{\mathrm{TB}}$ for atom pairs from edge features. By assembling all $\mathbf{H}_{ij}^{\rm TB}$ across the structure, one obtains a full TB Hamiltonian, which is then used to compute the electronic band structure. Therefore, a single forward pass provides both the interatomic potential and the full TB Hamiltonian. By formulating the problem within a multitask learning framework, UnIPH leverages shared representations between energy/force prediction and Hamiltonian learning. This leads to improved efficiency and consistency compared to training different tasks separately \cite{multitask}.

\subsection{The architecture of UnIPH}
\begin{figure}[h!]
\includegraphics[width=0.95\textwidth]{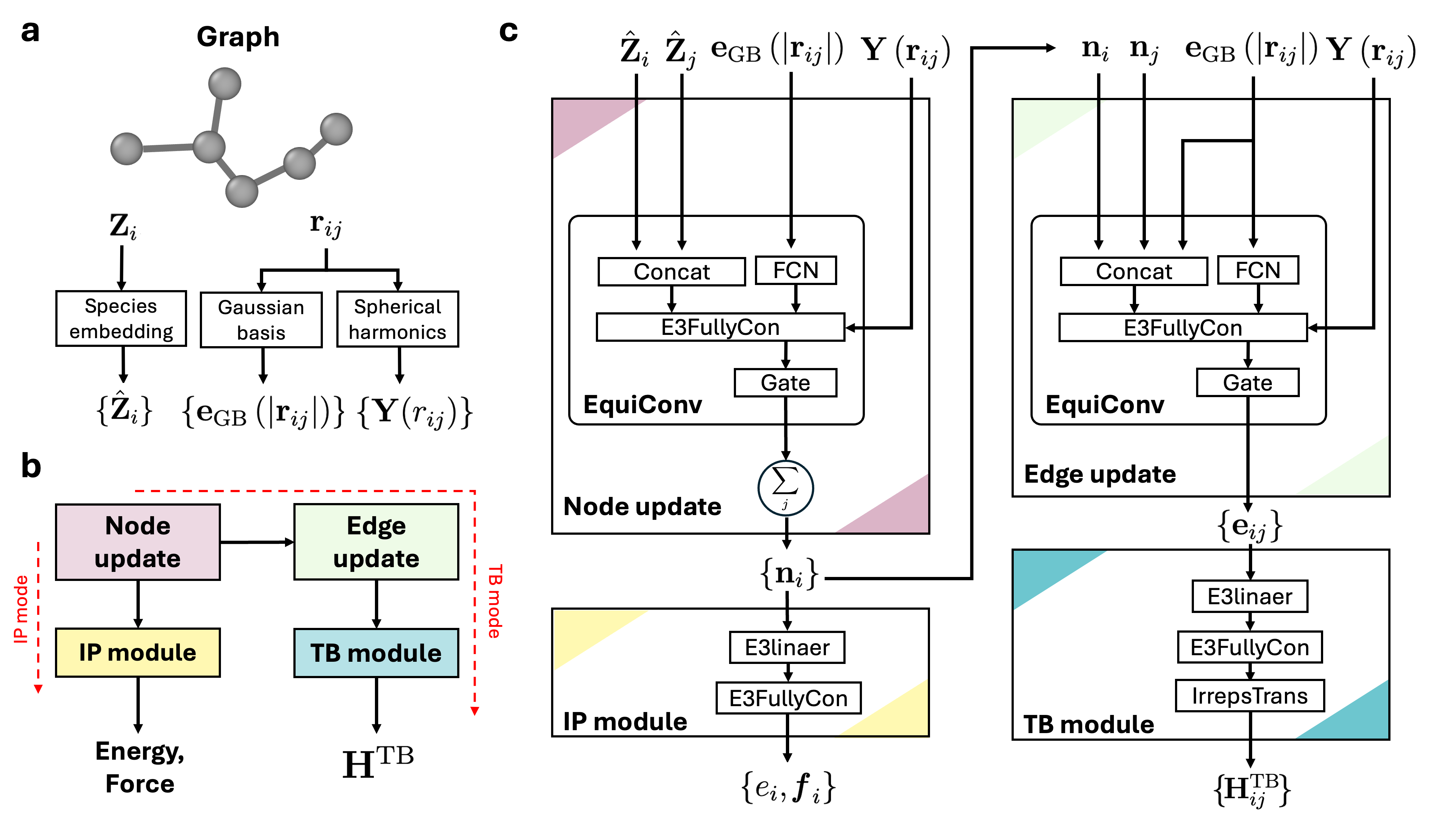}
\caption{The architecture of UnIPH.
(a) A graph is constructed from an atomic structure, with initial features based on atomic species and geometric information.
(b) Simplified overview of the UnIPH workflow, consisting of four modules: Node update (pink), Edge update (green), IP module (yellow), and TB module (blue). The workflow has two parallel pathways (denoted as IP mode and TB mode). The IP module predicts energy and forces and the TB module predicts TB Hamiltonians ($\mathbf{H}^{\rm TB}$).
(c) Schematic of detailed process within UnIPH. See details in \msec{sec:method:architecture}. \rev{Molecular structures in this figure were visualized using OVITO version 3.7.6.}}
\label{fig:architecture}
\end{figure}

The architecture of UnIPH is presented in \fig{fig:architecture}. As shown in \fig{fig:architecture}a, a graph is constructed from a small atomic configuration, where each atom $i$ is represented as a node. 
Edges are defined between neighboring atoms based on their relative displacement vectors $\mathbf{r}_{ij}$, determined within a cutoff distance $d_{\mathrm{cutoff}}$ and accounting for periodic boundary conditions (see \msec{sec:method:architecture} for details). Initial node features are determined by atom types and encoded using one-hot representations of the atomic numbers $\mathbf{Z}_i$ (e.g., H: $Z=1$, C: $Z=6$, Mo: $Z=42$).
These vectors are then embedded into a continuous feature space via a multilayer perceptron, producing $\mathbf{\hat{Z}}_i$, which serves as the initial node feature for the UnIPH forward pass.
Radial distance information is captured by expanding the magnitudes $\vert\mathbf{r}_{ij}\vert$ into a set of Gaussian radial basis functions (\eqn{eqn:method:gaussian} in \msec{sec:method:architecture}), resulting in $\mathbf{e}_{\mathrm{GB}}(\vert\mathbf{r}_{ij}\vert)$. Angular information is encoded using spherical harmonics $Y_{\ell m}(\hat{\mathbf{r}}_{ij})$, where $\ell$ is the angular momentum quantum number (degree) and $m$ is the magnetic quantum number (order), ranging from $-\ell$ to $\ell$. These functions describe the directionality of $\hat{\mathbf{r}}_{ij} = \mathbf{r}_{ij} / \vert\mathbf{r}_{ij}\vert$ in a manner equivariant under rotations and inversion.

\fig{fig:architecture}b presents a simplified schematic of the information flow in UnIPH, while the detailed processes are shown in \fig{fig:architecture}c. The node feature for each atom, $\mathbf{n}_i$, is computed in the Node Update block (pink). In this block, each node exchanges information with its neighbors through the EquiConv operator \cite{deephe3}. In EquiConv, the initial features of two connected atoms, $\mathbf{\hat{Z}}_i$ and $\mathbf{\hat{Z}}_j$, are concatenated and combined with the spherical harmonics $\mathbf{Y}(\mathbf{r}_{ij})$ via a fully connected tensor product (E3FullyCon). The weights for E3FullyCon are generated by applying a fully connected network (FCN) to the radial basis embedding $\mathbf{e}_{\mathrm{GB}}(\vert\mathbf{r}_{ij}\vert)$. After passing through a gating operation, the resulting messages are averaged over all neighbors $j$ of atom $i$, producing the updated node feature $\mathbf{n}_i$. The irreducible representations used for $\mathbf{n}_i$ are detailed in \msec{sec:method:architecture}.
The node feature $\mathbf{n}_i$ then propagates through two parallel pathways: the IP module (yellow) and the Edge Update module (green). In the IP module, $\mathbf{n}_i$ is first processed by an E3Linear layer (eqn. (12) in \cite{deephe3}), followed by a fully connected tensor product applied to itself. This produces a scalar atomic energy $e_i$, from which forces are obtained by differentiating $e_i$ with respect to atomic positions.
In the Edge Update module, the node feature $\mathbf{n}_i$ is used to compute the edge feature $\mathbf{e}_{ij}$. This process parallels the Node Update: $\mathbf{n}_i$ and $\mathbf{n}_j$ are concatenated and passed through an EquiConv layer to generate edge features for each atomic pair. The irreducible representations used for $\mathbf{e}_{ij}$ are detailed in \msec{sec:method:architecture}.
The resulting $\mathbf{e}_{ij}$ is then processed in the TB module. Here, $\mathbf{e}_{ij}$ is first transformed by an E3Linear layer and then by an E3FullyCon layer to produce a direct sum of irreducible representations of the TB Hamiltonian, $\mathbf{H}_{ij}^{\mathrm{1D}}$. Finally, the Clebsch–Gordan decomposition is applied in the IrrepsTrans operator to convert the direct sum of irreps in $\mathbf{H}_{ij}^{\mathrm{1D}}$ into the $N \times M$ TB Hamiltonian matrix, $\mathbf{H}_{ij}^{\rm TB}$, where $N$ and $M$ are the numbers of orbitals on atoms $i$ and $j$, respectively (see \msec{sec:method:architecture} for details).

An important feature of this architecture is the independence of the IP and TB pathways, allowing flexible usage. Either module can be selectively frozen during inference. For example, the trained UnIPH can be used solely for IP prediction or TB prediction, allowing for computational efficiency depending on the task. A benchmark comparison of the computational costs for TB and IP predictions across different IP and TB models is provided in Supplementary Section S2.

\subsection{Twisted bilayer graphene}\label{sec:TBG}
We first trained UnIPH on bilayer graphene to examine how a structural deformation influences the electronic structure of TBG. The training dataset contains 1,960 configurations generated under diverse conditions (interlayer shear, atomic perturbations, and ab initio finite-temperature molecular dynamics) computed using DFT and subsequently projected with Wannier90. Details of the configuration set are provided in \msec{sec:method:dataset}. \rev{A task-level comparison with the reference and baseline method families is provided in Supplementary Table~S1, and the sensitivity to the multitask loss weights is summarized in Supplementary Table~S2.} On the test set, UnIPH achieves \rev{mean absolute errors (MAEs) of 0.05~meV/atom for energy, 10.28~meV/\angstrom\ for the Cartesian force components, 11.34~meV/\angstrom\ for the force magnitudes, and 4.35~meV for TB Hamiltonian elements in bilayer graphene (Supplementary Figure~S1 and Supplementary Table~S3).} \rev{Repeated random-split evaluations and learning curves with respect to training epoch and training-set size are also reported (Supplementary Figures~S2 and S3 and Supplementary Table~S4).} \rev{To benchmark UnIPH against different existing approaches, conventional DFTB parameterizations and periodic GFN1-xTB were evaluated on the graphene test set (Supplementary Figure~S4 and Supplementary Table~S5), GRACE was fitted to the same graphene energy and force data (Supplementary Figures~S5 and S6 and Supplementary Tables~S6 and S7), and DeepH-E3 was trained on the same Wannier target and data split for a direct comparison of Hamiltonian accuracy (Supplementary Figures~S7 and S8 and Supplementary Table~S8) \cite{deephe3}. Operation-level timing comparisons are shown in Supplementary Figure~S9. Additional validation of Wannier-center displacements and symmetry-broken graphene environments is provided in Supplementary Figures~S10 and S11. The computational settings are reported in Methods sections \msecn{sec:method:dftb_xtb}, \msecn{sec:method:grace}, and \msecn{sec:method:deeph}.}

\begin{figure}[h!]
\includegraphics[width=0.95\textwidth]{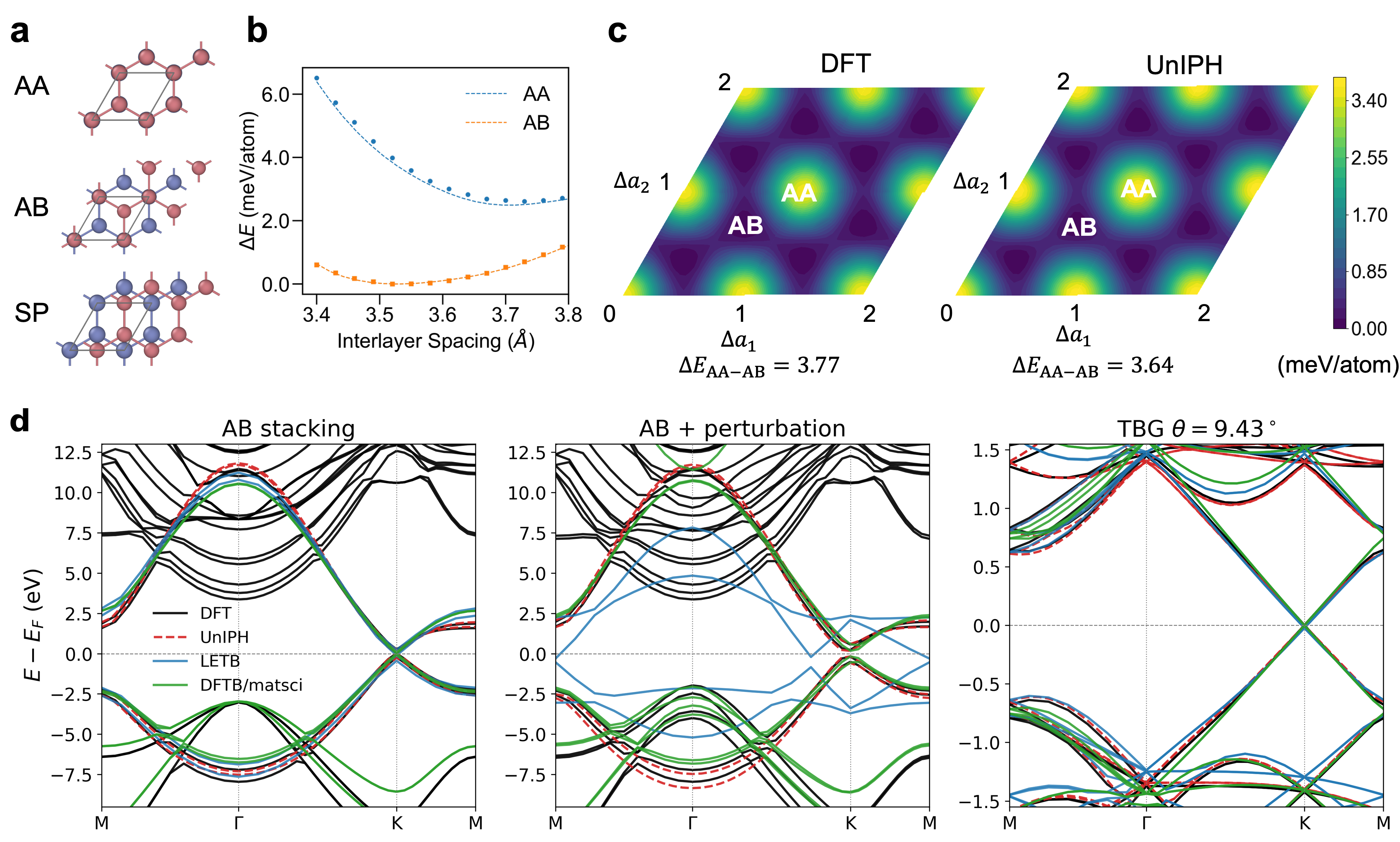}
\caption{Validation of UnIPH for IP and TB predictions in bilayer graphene.
(a) Three representative stacking configurations: AA, AB, and SP stackings. Red and blue spheres represent atoms in the top and bottom graphene layers, respectively.
(b) Energy evolution as a function of interlayer spacing for AA and AB stacking. DFT results (solid dots) are compared with UnIPH predictions (dotted lines). For both, the minimum energy of AB-stacking is set to zero for relative comparison.
(c) Generalized stacking fault energy (GSFE) for interlayer shear in bilayer graphene, obtained from DFT and UnIPH. Lateral displacements $\Delta a_1$ and $\Delta a_2$ are expressed in units of the lattice vectors $\mathbf{a}_1 = \sqrt{3}a\,\hat{\mathbf{x}}$ and $\mathbf{a}_2=a(\frac{\sqrt{3}}{2}\hat{\mathbf{x}}+\frac{3}{2}\hat{\mathbf{y}})$, where $a$ is the carbon--carbon bond length. The energy difference per atom between AA and AB stackings ($\Delta E_{\rm AA-AB}$) is also shown at the bottom.
\rev{d) Electronic band structures of pristine AB bilayer graphene, perturbed AB bilayer graphene, and TBG at $\theta=9.43^\circ$ (148 atoms) along M--$\Gamma$--K--M. DFT (black), UnIPH (red dashed), LETB (blue), and the representative DFTB/matsci baseline (green) are shown on the same $k$-path as the DFT reference.} \rev{Molecular structures in this figure were visualized using OVITO version 3.7.6.}}
\label{fig:TBG:IPTB}
\end{figure}

\fig{fig:TBG:IPTB} shows the performance of UnIPH as both IP and TB models for bilayer graphene. \fig{fig:TBG:IPTB}a illustrates three stacking configurations (AA, AB, and SP) whose relative energies govern a structural deformation during a TBG relaxation. In \fig{fig:TBG:IPTB}b, we compare the energy per atom obtained from DFT and UnIPH as a function of the interlayer spacing, with the minimum energy of AB stacking taken as the reference value (i.e., 0 meV/atom). The UnIPH model closely reproduces the DFT curve, capturing the interlayer energy profile. \fig{fig:TBG:IPTB}c presents GSFE calculations at a fixed interlayer distance of 3.5$\,\angstrom$. The UnIPH model accurately matches the GSFE landscape with the DFT calculations, including the AA–AB stacking energy difference (DFT: 3.77 meV/atom; UnIPH: 3.64 meV/atom). In addition, the predicted in-plane elastic constants ($c_{11}$ = 1123.6 GPa and $c_{12}$ = 155.8 GPa) from UnIPH agree well with DFT values ($c_{11}$ = 1060.2 GPa and $c_{12}$ = 123.6 GPa), confirming the fidelity of the model for both interlayer and intralayer energies.

\rev{Figure~\ref{fig:TBG:IPTB}d compares the band structures from DFT, UnIPH, LETB, and the representative DFTB/matsci baseline for pristine AB bilayer graphene, an atomically perturbed AB bilayer, and $9.43^\circ$ TBG. For pristine AB stacking and TBG, both UnIPH and LETB follow the DFT bands closely because both models were fitted to DFT-derived graphene data containing the corresponding stacking environments. For the atomically perturbed AB structure, UnIPH continues to reproduce the DFT bands, whereas LETB shows larger deviations because such atomic-scale perturbations are not represented in its fitting data and prescribed hopping form. DFTB/matsci captures the overall $\pi$-band dispersion across the three structures, indicating that its distance-dependent parameterization follows the general structural changes and therefore responds to the atomic perturbation, although the detailed band splittings show larger differences from DFT.}

\begin{figure}[h!]
\includegraphics[width=0.93\textwidth]{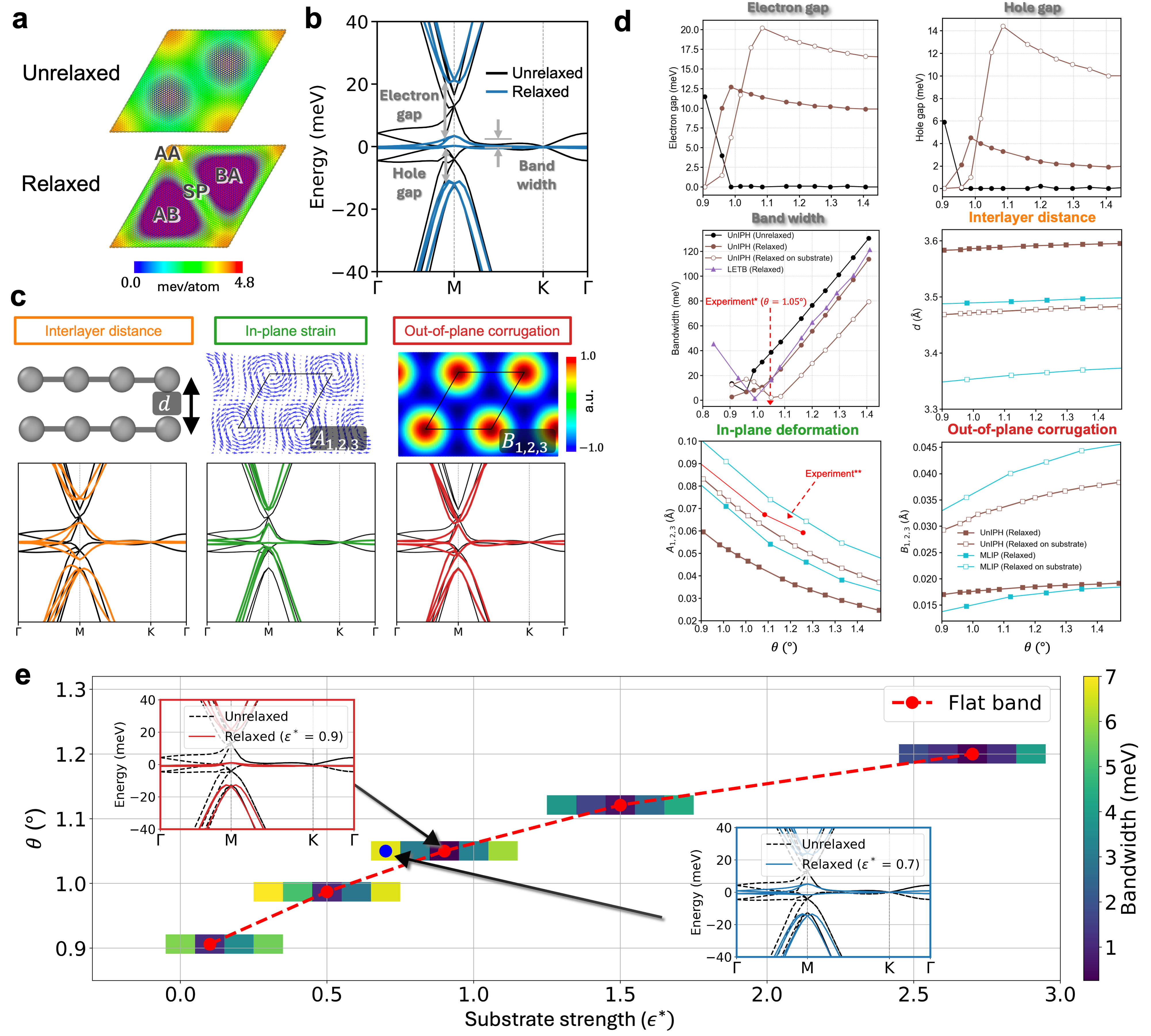}
\caption{The effect of structural deformation on electronic band structure of TBG.
(a) Energy maps of the unrelaxed and relaxed TBG at $1.05^\circ$, with stacking domains indicated.
(b) Electronic band structures of unrelaxed (black) and relaxed (blue) TBG predicted by UnIPH, with the electronic gap, hole gap, and bandwidth highlighted.
(c) Deformation modes in TBG: interlayer distance (orange), in-plane strain (green), and out-of-plane corrugation (red). The middle row represents each mode, with corresponding coefficients labeled as $d$, $A_{1,2,3}$, and $B_{1,2,3}$. The bottom row illustrates the isolated impact of each mode on the band structure of TBG: (i) $d$ only, (ii) $A_{1,2,3}$ only, and (iii) $B_{1,2,3}$ only. Deformation coefficients are taken from all-atom simulations of the $1.05^\circ$ TBG shown in (a) and (b).
(d) Evolution of band structure features (electronic gap, hole gap, and bandwidth) and deformation coefficients ($d$, $A_{1,2,3}$, and $B_{1,2,3}$) as a function of twist angle for unrelaxed, relaxed, and \rev{relaxed with the continuum-LJ mechanical interaction}. \rev{Experimental data are shown for comparison: Experiment$^*$ marks the experimentally observed magic angle ($\theta = 1.05^\circ$) \cite{cao2018}, and Experiment$^{**}$ denotes the in-plane strain coefficients $A_{1,2,3}$ obtained from electron diffraction \cite{choi2024,sung2022,yoo2019}.}
\rev{(e) Bandwidth map versus twist angle and the nominal LJ adhesion-strength scale ($\epsilon^*$), showing strain-mediated modulation within the mechanical proxy. The red dotted line marks the minimal-bandwidth trajectory. Insets show band structures at $\theta = 1.05^\circ$ for $\epsilon^* = 0.9$ (red) and $\epsilon^* = 0.7$ (blue).}}
\label{fig:TBG:RELAX}
\end{figure}

\fig{fig:TBG:RELAX} shows the effect of structural deformation on electronic band structure of TBG. \fig{fig:TBG:RELAX}a illustrates how the structural relaxation changes the energy landscape of TBG. \rev{We use the continuum LJ interaction as a computationally tractable mechanical boundary condition that avoids an explicit atomistic substrate and isolates the effect of prescribed substrate pulling. Si$_3$N$_4$ is used as the reference because it is a common experimental support for TBG and produces a stronger mechanical pulling effect than hBN with the reported LJ parameters \cite{choi2024}. For the substrate-supported cases, structural relaxation minimizes the sum of the UnIPH-predicted TBG energy and this external interaction. The substrate is not represented by atoms or electronic orbitals; after relaxation, the TBG Hamiltonian is evaluated from the carbon coordinates alone. This calculation therefore excludes direct charge transfer, orbital hybridization, dielectric screening, interface dipoles, and atomic-scale registry. Quantifying these direct electronic substrate effects would require calculating the electronic structure of the combined graphene--substrate system \cite{yang2011graphene}. Further details are provided in \msec{sec:method:tbg_lj}.}

Using UnIPH, we relax TBG at $\theta=1.05^\circ$ ($L = 13.4$ nm; 11,908 atoms) from an initial interlayer distance of 3.7$\,\angstrom$. Two energy maps in \fig{fig:TBG:RELAX}a reveal that the relaxation expands low-energy AB/BA stacking domains while shrinking a high-energy AA stacking region, forming sharp domain boundaries along SP stacking lines.
\fig{fig:TBG:RELAX}b shows the corresponding band structure evolution. In the band structure, the unrelaxed TBG exhibits negligible electron and hole gaps with a broad bandwidth. The relaxation opens finite gaps and narrows the bandwidth, resulting in an isolated flat band.

To examine the effects of different deformation types, the deformation in the relaxed structure from all-atom simulations is decomposed into three orthogonal modes: interlayer distance modulation, in-plane strain, and out-of-plane corrugation, following the procedure in \cite{choi2024}. The extracted parameters are the interlayer distance $d = 3.473\,\angstrom$, the amplitude of the first three in-plane strain coefficients $A_{1,2,3} = 0.067\,\angstrom$, and the amplitude of the first three out-of-plane corrugation coefficients $B_{1,2,3} = 0.033\,\angstrom$. The middle row of \fig{fig:TBG:RELAX}c provides a visualization of these deformation modes along with the corresponding parameter values.
The lower row of \fig{fig:TBG:RELAX}c illustrates the band structures obtained by applying each mode individually to the initial TBG structure ($\theta=1.05^\circ$).
As shown in \fig{fig:TBG:RELAX}c(i), reducing the interlayer spacing alone from $3.7$ to $3.473\,\angstrom$ produces isolated bands \cite{carr2018}, but the bandwidth remains larger than in the fully relaxed case (\fig{fig:TBG:RELAX}b). In-plane strain, which enlarges AB/BA regions while shrinking AA regions \cite{sung2022,zhang2018}, also leads to band isolation, but again with insufficient bandwidth reduction as shown in \fig{fig:TBG:RELAX}c(ii). Out-of-plane corrugation modulates a local interlayer spacing (increasing it in AA and decreasing it in AB/BA regions), thereby reducing interlayer hopping and partially narrowing the bandwidth \cite{lucignano2019,rakib2022}, but without opening band gaps as shown in \fig{fig:TBG:RELAX}c(iii).
These results highlight that no single deformation mode reproduces the full relaxation-induced isolated flat band: all three must be considered in combination to capture the complete electronic response of TBG. \rev{The three projected modes qualitatively reproduce the band structure of the fully relaxed TBG, indicating that they capture the dominant deformations without higher-order basis functions; higher-order in-plane and out-of-plane bases are described in Ref.~\cite{choi2024}.}

\fig{fig:TBG:RELAX}d shows the evolution of the band structure and structural deformation in TBG relaxation across a range of twist angles ($\theta$). Electron gap, hole gap, and bandwidth are presented as key band structure characteristics, computed for three different cases: unrelaxed, relaxed, and relaxed on the substrate. For the bandwidth, we also add the result from the LETB model \cite{letb}, which corresponds to the relaxed TBG.
The electron and hole gaps display similar trends with a twist angle change. In both the relaxed with/without substrate cases, finite gaps appear at higher twist angles, whereas unrelaxed TBG remains gapless in this regime. As $\theta$ decreases, the gaps increase, peaking at $0.987^\circ$ for the relaxed TBG and $1.085^\circ$ for the relaxed TBG on substrate, before closing below $0.906^\circ$. In contrast, the unrelaxed case stays gapless for $\theta > 0.987^\circ$, with gaps emerging only at smaller twist angles.
The bandwidth, an important feature related to the emergence of superconductivity, monotonically decreases with decreasing twist angle in all cases. The minimum bandwidth (flat band) occurs at different twist angles depending on the case: $1.05^\circ$ for the relaxed TBG on the substrate, $0.987^\circ$ for the relaxed TBG, $0.906^\circ$ for the unrelaxed TBG, and $0.987^\circ$ for LETB. \rev{Within this mechanical proxy, proximity of the $1.05^\circ$ minimum to the experimental magic angle does not show that the complete substrate interface is reproduced, because direct electronic and dielectric effects are absent \cite{cao2018}.} 

Three key structural deformation parameters, interlayer distance ($d$), in-plane strain coefficients ($A_{1,2,3}$), and out-of-plane corrugation coefficients ($B_{1,2,3}$) are also plotted in \fig{fig:TBG:RELAX}d across twist angles. Because these parameters are zero for the unrelaxed TBG (i.e., no deformation), we focus on the relaxed and relaxed-on-the-substrate cases obtained from our UnIPH model. For comparison, we also plot results from MLIP (hybrid neural network potential) \cite{hnn}.
The interlayer distance $d$ remains relatively constant across twist angles within each relaxation scheme; however, the effect of the substrate is clearly significant. In the UnIPH predictions, the average interlayer spacing decreases from 3.59 $\angstrom$ (relaxed) to 3.48 $\angstrom$ (relaxed on the substrate), while in the MLIP results, $d$ decreases from 3.49 $\angstrom$ to 3.35 $\angstrom$ with the substrate.
The in-plane strain coefficients ($A_{1,2,3}$) increase monotonically as the twist angle decreases. The presence of the substrate enhances these coefficients at any given twist angle in both UnIPH and MLIP. This trend arises because the reduced interlayer spacing due to the substrate effect increases the energy difference between AB and AA stackings, thereby increasing in-plane strain to minimize the total energy despite the associated in-plane strain energy. Computed $A_{1,2,3}$ values from the electron diffraction experiment are also included for validation and comparison \cite{choi2024,sung2022,yoo2019}.
The out-of-plane corrugation coefficient ($B_{1,2,3}$) tends to decrease as the twist angle decreases. Similar to the $A_{1,2,3}$, the substrate effect increases $B_{1,2,3}$ for both UnIPH and MLIP simulations. This is again attributed to the enhanced stacking energy contrast, which allows for greater corrugation to reduce total energy at the cost of out-of-plane strain. However, overall the magnitude of $B_{1,2,3}$ is smaller than that of $A_{1,2,3}$, indicating that in-plane deformation contributes more in the relaxation of TBG. 

\fig{fig:TBG:RELAX}e examines how varying \rev{the prescribed mechanical interaction} influences both structural relaxation and electronic properties of TBG. \rev{Because the continuum substrate has no electronic degrees of freedom in this model, all changes in electronic properties arise solely from substrate-induced changes in the relaxed TBG geometry.} In these simulations, only the LJ potential energy parameter ($\epsilon$) is scaled, while the distance parameter ($\sigma$) is fixed. The dimensionless parameter $\epsilon^*$ \rev{is a nominal LJ adhesion-strength scale factor}: $\epsilon^* = 1.0$ \rev{uses the stated Si$_3$N$_4$ reference parameters} and $\epsilon^* = 0$ \rev{is freestanding TBG; it is not a material label or a descriptor of direct substrate electronic effects}.
By tuning $\epsilon^*$, we investigate how the substrate-induced pulling influences the relaxed TBG structure and its corresponding electronic band structure. The heatmap in \fig{fig:TBG:RELAX}e shows the computed bandwidth as a function of twist angle and $\epsilon^*$. Dark blue regions indicate a near-zero bandwidth, corresponding to the formation of an isolated flat band, while yellow regions indicate larger bandwidths. The red dotted curve traces the near-zero bandwidth (flat band) across twist angles and \rev{the prescribed mechanical parameters}. This trend reveals that achieving flat bands at larger twist angles requires \rev{a stronger prescribed mechanical interaction}; for instance, at $\theta = 1.2^\circ$, an $\epsilon^*$ of approximately 2.7 is needed to approach a near-zero bandwidth.
Insets in \fig{fig:TBG:RELAX}e illustrate band structures at a fixed twist angle of $1.05^\circ$ under different \rev{mechanical-interaction strengths} ($\epsilon^*$). The band structure in the red box, corresponding to the red point (relaxed with $\epsilon^*=0.9$), exhibits an isolated flat band, while the band structure in the blue box corresponds to the blue point (relaxed with $\epsilon^*=0.7$) and shows a larger bandwidth. These results demonstrate that \rev{the prescribed mechanical interaction} modulates band structure by altering the deformation during TBG relaxation, thereby enabling the formation of an isolated flat band even at lower or higher values relative to the conventional magic angle ($\approx1.1^\circ$).

\subsection{Monolayer MoS\texorpdfstring{\textsubscript{2}}{2}}\label{sec:MoS2}
We trained UnIPH on monolayer MoS$_2$ (2H phase) to investigate how structural deformation affects its electronic structure. The training dataset is composed of two types of configurations: (1) strained and perturbed $1\times1$ monolayer MoS$_2$ (200 data points) and (2) ab initio molecular dynamics simulations of a $4\times4$ monolayer MoS$_2$ supercell at 50, 100, 200, and 300~K (400 data points), both computed using DFT and subsequently projected with Wannier90.
Details of the configuration set are provided in \msec{sec:method:dataset}. On the test set, UnIPH achieves \rev{MAEs of 0.32~meV/atom for energy, 18.93~meV/\angstrom\ for the Cartesian force components, 19.73~meV/\angstrom\ for the force magnitudes, and 1.86~meV for TB Hamiltonian elements in monolayer MoS$_2$}, with quantitative comparisons reported in Supplementary Section S1. \rev{The corresponding repeated random-split evaluations and learning curves are reported in Supplementary Section S1.}

\fig{fig:MOS2:IPTB} validates the trained UnIPH for the monolayer MoS$_2$.
\fig{fig:MOS2:IPTB}a compares the phonon dispersion relations predicted by UnIPH with DFT results along the high-symmetry path $\Gamma$–M–K–$\Gamma$ of the hexagonal Brillouin zone. The close agreement shows that UnIPH accurately captures the elastic and vibrational properties of the monolayer MoS$_2$.  
\fig{fig:MOS2:IPTB}b shows the electronic band structures obtained from UnIPH and DFT for both the unstrained and uniformly strained monolayer MoS$_2$ unit cell, with strain components $\varepsilon_{xx} = \varepsilon_{yy} = \varepsilon_{xy} = 0.015$. The strong agreement in both cases confirms the reliability of UnIPH. 
\fig{fig:MOS2:IPTB}c examines the band gap evolution under the uniform strain. The left panel illustrates the monolayer structure and the applied strain tensor with three independent components ($\varepsilon_{xx}$, $\varepsilon_{yy}$, and $\varepsilon_{xy}$). Each strained configuration is fully relaxed before the band structure calculation. The five right panels show band gap variations as a function of biaxial strain ($\varepsilon_{xx}$ and $\varepsilon_{yy}$) for fixed shear strains $\varepsilon_{xy} = -0.02$, $-0.01$, $0$, $0.01$, and $0.02$. Overall, a compressive strain increases the band gap, while tensile strain reduces it; this trend holds for all $\varepsilon_{xy}$ values, with shear strain further modulating both the magnitude and the direct/indirect nature of the gap. For comparison, results for strained but unrelaxed structures are shown in Supplementary Figure~S12, highlighting that the relaxation is critical in determining the direct or indirect character of the gap.

Using UnIPH, we next examine a substrate-driven strain-engineering problem in monolayer MoS$_2$ (\fig{fig:MOS2:LDOS}; details in \msec{sec:method:mos2_substrate}). A $20\times20$ supercell is placed on a non-uniform continuum substrate feature and structurally relaxed for substrate amplitudes $A=0$, 1, and 2~\angstrom. \fig{fig:MOS2:LDOS}a shows that the reference case $A=0$ remains nearly flat, whereas increasing substrate amplitude generates a localized deformation centered on the feature. The associated maximum principal strain rises from essentially zero to 0.209\% at $A=1$~\angstrom\ and 0.903\% at $A=2$~\angstrom. \rev{The continuum substrate enters only through mechanical relaxation. The Hamiltonian and LDOS are subsequently computed from the relaxed MoS$_2$ coordinates without substrate atoms or orbitals. \fig{fig:MOS2:LDOS} therefore reports the strain-mediated band-edge response conditional on the prescribed continuum interaction, not the complete response of a specific MoS$_2$--substrate interface.}

\fig{fig:MOS2:LDOS}b first shows the corresponding local band-gap maps. The $A=0$ case is spatially uniform, with a local band gap of 1.4407 eV. Once the substrate height becomes nonzero, a localized gap minimum appears near the strain maximum and deepens with increasing amplitude, reaching 1.4317 eV for $A=1$\,\angstrom\ and 1.3537 eV for $A=2$\,\angstrom. These correspond to local gap reductions of 9.0 and 87.0 meV, respectively. \rev{Within this mechanical proxy, the localized deformation modifies the learned MoS$_2$ Hamiltonian and produces a spatially varying band-edge landscape.}

Below the local band-gap maps, the diagonal line cuts resolve how the band edges form this potential. The gap narrowing is not a rigid spectral shift: at $A=2$\,\angstrom, the CBM (conduction band minimum) moves downward by about 58.5 meV, while the VBM (valence band maximum) shifts upward by about 28.5 meV, producing a localized band-edge pocket centered on the strained region.
This trend is consistent with prior MoS$_2$ strain-engineering studies in which tensile strain reduces the gap and patterned strain creates spatially varying electronic landscapes \cite{conley2013,castellanos2013,hui2014,li2015}. 
This type of local band-edge landscape could be useful for designing quantum-emitter architectures and other strain-engineered optoelectronic functionalities in TMD monolayers \cite{peng2020}. Such analysis requires large-scale structural relaxation together with deformation-dependent electronic Hamiltonian predictions, which are enabled by UnIPH. Direct DFT calculations for these large, spatially nonuniform strained systems would be computationally prohibitive, while classical atomistic simulations do not provide access to the corresponding electronic structure.

\section{Discussion}\label{sec:discussion}
In this work, we introduce UnIPH, a graph neural network for unified learning of interatomic potentials and TB Hamiltonians for a large-scale atomic structure. Built on an equivariant graph neural network framework, UnIPH is trained on a dataset comprising total energies and atomic forces (as node targets) and TB Hamiltonians (as edge targets), all obtained from DFT calculations and Wannier projections.
We first train UnIPH on bilayer graphene to capture structural relaxation and electronic reconstruction in TBG over a range of twist angles. The simulations reveal how distinct deformation modes (interlayer distance variation, in-plane strain, and out-of-plane corrugation) govern the emergence of isolated flat bands. \rev{We further show, using a continuum-LJ mechanical proxy, that prescribed substrate coupling can modulate these deformations and generate isolated flat bands at various angles. Within this proxy, the electronic changes arise solely from the substrate-induced changes in relaxed geometry.}
Beyond bilayer graphene, UnIPH trained on monolayer MoS$_2$ reproduces the phonon dispersion and strain-dependent band gap evolution with DFT-level accuracy. \rev{Using a patterned continuum-LJ mechanical proxy for a large supercell, the model predicts localized strain pockets and the resulting spatial modulation of the CBM, VBM, and local band gap.}
\rev{Possible future research directions include:}
\begin{itemize}
    \item \rev{\textbf{Extension to other material and structural classes}: Potential extensions include twisted multilayers, heterostructures, and bulk or lower-symmetry systems. Such applications require system-specific DFT+Wannier training data, a consistent localized-orbital basis across the relevant configurations, and separate validation; they therefore remain future work.}

    \item \rev{\textbf{Molecular dynamics simulations and phonon–electron interactions}: Joint prediction of atomic forces, total energies, and TB Hamiltonians may support future studies of phonon--electron coupling in extended systems. Such applications will require dedicated dynamical validation.}
\end{itemize}
\rev{Altogether, UnIPH provides a unified workflow for the separately trained bilayer-graphene and monolayer-MoS$_2$ models studied here. We anticipate that the unified formulation can be extended to other material and structural classes; however, each extension will require system-specific training data and validation.}

\section{Methods}\label{sec:method}
\subsection{DFT calculations}\label{sec:method:dft}
All DFT calculations were performed using the plane-wave pseudopotential code Quantum ESPRESSO \cite{qe}. The exchange–correlation energy was computed by the generalized gradient approximation using the Perdew–Burke–Ernzerhof functional. A kinetic energy cutoff of 50 Ry was used for the plane-wave basis set, with charge density cutoffs set to 400 Ry for bilayer graphene and 500 Ry for monolayer MoS$_2$. The Brillouin zone was sampled using a $\Gamma$-centered Monkhorst–Pack grid, with sufficient density to ensure total energy convergence better than 0.001 meV/atom for all systems. To eliminate spurious interactions between periodic images, all structures were placed in supercells with at least 20 $\angstrom$ of vacuum in the out-of-plane direction. Self-consistent field calculations were converged to an energy threshold of $10^{-7}$ Ry. For bilayer graphene, van der Waals interactions were included using the Grimme D3 dispersion correction scheme to accurately account for interlayer coupling.

\subsection{Wannier90 process}\label{sec:method:wannier}
Maximally localized Wannier functions (MLWFs) were constructed using the Wannier90 code to obtain tight-binding (TB) Hamiltonians in a localized basis. For each system, additional non-self-consistent field DFT calculations were performed on a dense $k$-mesh, and the resulting wavefunctions were used for Wannier projections.  
In bilayer graphene, MLWFs were initialized from the $p_z$ orbital on each carbon atom to capture the $\pi$-band character~\cite{letb}. A frozen window of [–1.2, 1.5]~eV and a disentanglement window extending up to 17~eV were applied to isolate the low-energy $\pi$ bands while maintaining numerical stability.  
For monolayer MoS$_2$, MLWFs were initialized from Mo $d$ and S $p$ orbitals, with a disentanglement window of [–10, 2.5]~eV. 
\rev{Because the Wannier basis is orthonormal, $\mathbf{S}=\mathbf{I}$ and no separate overlap matrix needs to be calculated.}
\rev{To monitor the reliability and consistency of the localized basis, we tracked the spatial offset between atomic sites and their associated Wannier centers (e.g., $\delta_{C-{p_{\rm z}}}$ for carbon). During dataset preparation, the 0.01~\AA{} value was used as a conservative empirical diagnostic. A comparison of the archived Wannierized calculations with the released datasets showed that 98 of 2,058 bilayer-graphene candidates (4.76\%) were not included in the released 1,960-configuration dataset, whereas all 600 archived Wannierized MoS$_2$ configurations were included. Center displacement alone does not distinguish Wannierization failure from a converged, accurate Hamiltonian; the relevant requirement for learning is that the orbital assignment and Hamiltonian elements vary consistently across configurations. Further details are provided in \msec{sec:method:dataset} and Supplementary Section S3.}

\rev{As representation-level checks, we further applied the atom-centered $p_z$ Wannier construction to a DFT-relaxed reconstructed graphene divacancy and a hydrogen-terminated zigzag graphene nanoribbon. The mean (maximum) atom--center displacements were 0.00832 (0.03759)~\AA{} for the divacancy and 0.0155 (0.1018)~\AA{} for the nanoribbon. Despite these shifts, the Wannier-interpolated bands closely matched the corresponding DFT bands within the frozen window (Supplementary Section S3). These calculations test the localized reference construction only; neither structure was used to train or evaluate UnIPH.}

\subsection{UnIPH architecture and equivariant representation}\label{sec:method:architecture}
Some details of UnIPH are described here. To encode interatomic distance information, a set of Gaussian basis functions is used to expand the scalar distance between atom pairs. Each edge feature is defined as
\begin{equation}\label{eqn:method:gaussian}
    e_{\rm{GB}}(|\mathbf{r}_{ij}|)_n = \exp\left[ -\frac{(|\mathbf{r}_{ij}| - r_n)^2}{2\Delta^2} \right],
\end{equation}
where $r_n$ denotes the center of the $n$-th Gaussian function, where $n = 1, \dots, N$. The $\Delta$ controls the width (standard deviation) of each Gaussian.
In the implementation, the radial embedding is generated with Gaussian \texttt{soft\_one\_hot\_linspace} features from 0 to the cutoff radius and multiplied by $\sqrt{N}$; $N=14$ for bilayer graphene and $N=48$ for monolayer MoS$_2$.

\tab{tab:method:gnn_parameter} summarizes the parameters used for constructing the graph neural network for each material system, including the cutoff distance, the maximum angular momentum $\ell_{\max}$, and the specific irreducible representations (irreps) assigned to node and edge features. For both systems, the network uses one equivariant message-passing layer. The listed $\ell_{\max}$ controls the spherical-harmonic edge attributes, while the intermediate node and edge channels are those shown in the irreps column.

\rev{UnIPH predicts a Hamiltonian block for every directed graph edge $(i,j,\mathbf{R})$ within the cutoff, including the zero-shift self-loop $(i,i,\mathbf{0})$ that represents the on-site block; no separate node-level on-site branch is used, and the node-level output serves only the energy and force predictions. The blocks are not strictly two-center. The node update first aggregates the neighboring atoms, and the edge update then combines the updated features of both endpoint atoms with the edge geometry, so the edge feature that becomes a Hamiltonian block already encodes the many-body local environment rather than only the pair identity and separation. With the single message-passing layer used here, this environmental dependence is bounded by the union of the one-hop neighborhoods of the two endpoint atoms, and no direct hopping block is generated beyond the graph cutoff, which is 5.0~\angstrom\ for bilayer graphene and 7.0~\angstrom\ for monolayer MoS$_2$.}

\subsubsection{Conversion between tensor product and direct sum of irreducible representations} 
The tensor product of two irreducible representations (irreps) labeled by angular momenta $l_1$ and $l_2$ decomposes into a direct sum of irreps according to the following:
\begin{equation}
l_1 \otimes l_2 = \lvert l_1 - l_2 \rvert \oplus \cdots \oplus (l_1 + l_2).
\end{equation}
For example, when $l_1 = l_2 = 1$, the decomposition results in:
\begin{equation}
1 \otimes 1 = 0 \oplus 1 \oplus 2.
\end{equation}
This indicates that a $3\times 3$ tight-binding (TB) Hamiltonian defined on the $p-p$ orbital interaction can be decomposed into invariant subspaces corresponding to a direct sum of angular momenta $l = 0, 1,$ and $2$.
In equivariant graph neural networks, such direct sums are used to represent features of either edge or node. The transformation between the tensor product basis and the irreducible direct sum basis (and vice versa) is carried out using Clebsch–Gordan (CG) coefficients.
Given two input features $\mathbf{x}^{(l_1)} \in \mathbb{C}^{2l_1 + 1}$ and $\mathbf{x}^{(l_2)} \in \mathbb{C}^{2l_2 + 1}$, their tensor product is projected onto an irrep of angular momentum $l$ as:
\begin{equation}
\hat{x}^{(l)}_m = \sum_{m_1 = -l_1}^{l_1} \sum_{m_2 = -l_2}^{l_2} C^{l m}_{l_1 m_1,\, l_2 m_2} \cdot x^{(l_1)}_{m_1} \cdot x^{(l_2)}_{m_2},
\end{equation}
where  $\hat{x}^{(l)}_m \in \mathbb{C}$ is the component of the output feature that transforms under the irrep $l$ with magnetic quantum number $m$, and $C^{l m}_{l_1 m_1,\, l_2 m_2} \in \mathbb{R}$ is the corresponding CG coefficient.
The inverse CG transformation reconstructs the coupled tensor product basis from the irreducible components. Given the projected features $\hat{x}^{(l)}_m$, the original tensor product can be recovered via:
\begin{equation}\label{eqn:method:reCG}
x^{(l_1)}_{m_1} \cdot x^{(l_2)}_{m_2} = \sum_{l = |l_1 - l_2|}^{l_1 + l_2} \sum_{m = -l}^{l} C^{l m}_{l_1 m_1,\, l_2 m_2} \cdot \hat{x}^{(l)}_m.
\end{equation}
This inverse mapping expands the irreducible form into the full tensor product basis using the CG coefficients. In our framework, this operation (labeled as IrrepsTrans in \rev{Figure 2}) is used to transform edge features back into TB Hamiltonians.
\rev{After assembling the predicted Bloch matrix $\widetilde{\mathbf{H}}$, we apply $\mathbf{H}=(\widetilde{\mathbf{H}}+\widetilde{\mathbf{H}}^{\dagger})/2$ before diagonalization, so Hermiticity is enforced exactly. Because the two directions of a pair are predicted independently, the constraint is imposed at this assembly stage rather than by the neural-network architecture.}

\subsection{Training}\label{sec:method:training}
UnIPH is trained by minimizing a weighted sum of mean squared errors (MSEs) over three target properties: energy, force, and TB Hamiltonian elements. To ensure comparable scales across targets, the training targets are stored in normalized units and converted back to physical units for evaluation using the constants below.
First, the total energy from density functional theory (DFT) is expressed per atom, mean-centered across the dataset, and scaled by the root-mean-square (RMS) magnitude of atomic forces in the training set ($\sigma_f$):
\begin{equation}
\bar{E} = \frac{E/n_{\mathrm{atom}} - \langle E \rangle}{\sigma_f}.
\end{equation}
Similarly, atomic forces are normalized by the same factor $\sigma_f$:
\begin{equation}
\bar{\mathbf{f}}_i = \frac{\mathbf{f}_i}{\sigma_f}.
\end{equation}
Each element of the TB Hamiltonian is normalized by the standard deviation of all TB matrix elements in the dataset ($\sigma_{\mathrm{TB}}$):
\begin{equation}
\mathbf{\bar{H}}^{\mathrm{TB}}_{ij} = \frac{\mathbf{H}^{\mathrm{TB}}_{ij}}{\sigma_{\mathrm{TB}}}.
\end{equation}
\rev{The normalization constants used for the two material systems are listed in \tab{tab:method:normalization_values}.}

The total loss function used for training is defined as a weighted sum of normalized MSE losses over the energy, forces, and TB Hamiltonian predictions:
\begin{equation}
\mathcal{L}_{\mathrm{total}} = \frac{1}{N_{\mathrm{data}}} \sum_{n=1}^{N_{\mathrm{data}}} \left[
w_E \cdot \mathcal{L}_E^{(n)} + w_F \cdot \mathcal{L}_F^{(n)} + w_{\mathrm{TB}} \cdot \mathcal{L}_{\mathrm{TB}}^{(n)}
\right],
\end{equation}
where each $n$ component is defined as follows:
\begin{align}
    \mathcal{L}_E &= \left( \bar{E}^{\mathrm{pred}}_n - \bar{E}^{\mathrm{true}}_n \right)^2 \\
    \mathcal{L}_F &= \frac{1}{3N_{\mathrm{atoms}}} \sum_{i=1}^{N_{\mathrm{atoms}}} \sum_{\alpha=x,y,z} \left( \bar{f}^{\mathrm{pred}}_{i\alpha} - \bar{f}^{\mathrm{true}}_{i\alpha} \right)^2 \\
    \mathcal{L}_{\mathrm{TB}} &= \frac{1}{N_{\mathrm{edge}}} \sum_{(i,j)} \left\| \mathbf{\bar{H}}^{\mathrm{TB,pred}}_{ij} - \mathbf{\bar{H}}^{\mathrm{TB,true}}_{ij} \right\|_F^2
\end{align}
where $\|\cdot\|_F$ denotes the Frobenius norm, and $N_{\rm data}$ is the total number of training data, $N_{\rm atoms}$ is the total number of atoms in data indexed by $n$, and $N_{\rm edge}$ is the number of edges in data set.
For monolayer MoS$_2$, which contains multiple orbital-pair edge types, the TB loss is computed separately for each edge type present in the batch and then averaged over those edge types. Here ``pred'' and ``true'' denote UnIPH predictions and DFT+Wannier90 reference values, respectively.
\rev{The weight coefficients $w_E$, $w_F$, and $w_{\mathrm{TB}}$ multiply losses defined in normalized target spaces and are used to balance the three learning objectives. The reported weight coefficients were chosen empirically for this purpose rather than through a hyperparameter search. For bilayer graphene, we use $w_E = 500$, $w_F = 1$, and $w_{\mathrm{TB}} = 50$; for monolayer MoS$_2$, we use $w_E = 100$, $w_F = 10$, and $w_{\mathrm{TB}} = 1$.}

\rev{We use a local sensitivity protocol in which the data split and model-initialization seed are fixed at seed 0 and the energy, force, or Hamiltonian weight coefficient is doubled while the other two weight coefficients, model architecture, optimizer, and training protocol are held unchanged.}

\subsection{DFT dataset configurations}\label{sec:method:dataset}
\rev{We compared the archived Wannierized configurations with the released datasets. For bilayer graphene, 1,960 of 2,058 archived candidates were included, while 98 (4.76\%) were not included. All 98 of these configurations had a configuration-averaged atom--center displacement above 0.01~\AA{}. For monolayer MoS$_2$, all 600 archived Wannierized configurations were included in the released dataset, so no configurations were omitted.}
The atomic configurations in \tab{tab:method:dft_dataset} were generated by systematically varying atomic positions through in-plane strain, interlayer shifts, thickness modulation, and ab initio molecular dynamics (AIMD) at finite temperature. In-plane strain was introduced via the strain tensor
\begin{equation}
    \mathbf{R}' = (\mathbf{I} + \boldsymbol{\varepsilon}_{\rm 2D}) \, \mathbf{R},
\end{equation}
where $\mathbf{R}$ and $\mathbf{R}'$ are the original and strained lattice vectors, respectively, and $\boldsymbol{\varepsilon}_{\rm 2D}$ is the applied two-dimensional strain tensor,
\begin{equation}
\boldsymbol{\varepsilon}_{\mathrm{2D}} =
\begin{bmatrix}
\varepsilon_{xx} & \varepsilon_{xy} & 0 \\
\varepsilon_{xy} & \varepsilon_{yy} & 0 \\
0 & 0 & 0
\end{bmatrix}.
\end{equation}
Random atomic displacements were introduced to mimic thermal fluctuations and structural perturbations.
All configurations were computed using DFT, and the resulting electronic structures were projected onto localized orbital bases using Wannier90 to construct TB Hamiltonians.
For bilayer graphene, a $p_z$-like Wannier orbital was assigned to each carbon atom, yielding scalar TB Hamiltonian elements representing $p_z$–$p_z$ interactions.
For monolayer MoS$_2$, the basis set comprised $d$-like (Mo) and $p$-like (S) Wannier orbitals, producing $3\times 3$, $5\times 5$, and $3\times 5$ (and $5\times 3$) TB Hamiltonian blocks for Mo–Mo, S–S, and Mo–S couplings, respectively.
\rev{The orbital ordering and identity were held fixed across configurations. For each material, the same species-specific initial projections and orbital order were used for every configuration, and atom indices were preserved within each structure series. For graphene, the Wannier orbital index corresponded directly to the carbon atom index because each atom carried one $p_z$-like orbital. For MoS$_2$, the Wannier orbital indices were grouped into fixed blocks associated with each atom index, comprising five $d$-like orbitals for each Mo atom and three $p$-like orbitals for each S atom. No additional orbital permutation or sign/phase alignment was applied between configurations.}

\subsection{Evaluation metrics}\label{sec:method:metrics}
\rev{Because the force term of the training objective acts on Cartesian force components whereas the parity plots above compare force magnitudes, we report both error metrics explicitly. They are defined as}
\begin{equation}
\label{eq:method:force_metrics}
\rev{
\mathrm{MAE}_{\mathrm{comp}} = \frac{1}{3N}\sum_{i=1}^{N}\sum_{\alpha=x,y,z}
\left| F^{\mathrm{pred}}_{i\alpha} - F^{\mathrm{ref}}_{i\alpha} \right|,
\qquad
\mathrm{MAE}_{\mathrm{mag}} = \frac{1}{N}\sum_{i=1}^{N}
\left| \lVert \mathbf{F}^{\mathrm{pred}}_{i} \rVert_2 - \lVert \mathbf{F}^{\mathrm{ref}}_{i} \rVert_2 \right|,
}
\end{equation}
\rev{where $N$ is the number of atoms in the test set and $F_{i\alpha}$ are forces in physical units. Both were evaluated and listed in Supplementary Table~S3. All remaining comparisons in the Supplementary Information, namely the random-split replicates, the learning curves, and the DFTB and GRACE benchmarks, report $\mathrm{MAE}_{\mathrm{comp}}$ and are labelled accordingly in their respective tables.}

\subsection{Repeated-split and learning-curve protocol}\label{sec:method:learning_curves}
\rev{We performed two distinct analyses. First, to assess sensitivity to data selection and model initialization, we trained each material on the complete training pool in three end-to-end replicates, varying the frame-level random split and model-initialization seed together. Each replicate used an independently generated 90/5/5 training/validation/test split. Second, to construct the training-set-size learning curves, we held the validation and test partitions fixed within each replicate and fitted models to nested subsets containing 20, 40, 60, 80, and 100\% of that replicate's training pool. The corresponding training-set sizes were 353, 706, 1,058, 1,411, and 1,764 configurations for bilayer graphene and 108, 216, 324, 432, and 540 configurations for monolayer MoS$_2$.}

\rev{All fractions used the same model architecture, loss definition, and optimization protocol described in \msec{sec:method:training}. Bilayer graphene was trained for at most 200 epochs with an initial learning rate of $5\times10^{-4}$, and monolayer MoS$_2$ was trained for at most 100 epochs with an initial learning rate of $1\times10^{-3}$. All runs used batch size 1. The checkpoint with the lowest validation total loss was selected, and the test partition was evaluated only after checkpoint selection.}

\subsection{DFTB and GFN1-xTB benchmark}\label{sec:method:dftb_xtb}
\rev{The benchmark contains 98 configurations, 1,696 atoms, and 5,088 Cartesian force components from the graphene test split. The energy label is an extensive quantity and was divided by the number of atoms before comparison. For each method, the mean test-set energy per atom was subtracted to remove the method-dependent absolute energy zero. Force-component errors were evaluated over all Cartesian components with atom weighting. No training normalization factor was applied to the physical force labels.}

\rev{The comparison was restricted to graphene. Raw Hamiltonian elements were not compared: UnIPH uses an orthonormal $p_z$ Wannier basis, whereas DFTB/xTB use nonorthogonal $2s/2p$ atomic-orbital bases with an explicit overlap matrix. We therefore compare relative energies and forces for all methods and band structures for DFTB/matsci, using observables that are common to the models being compared.}

\rev{All DFTB and periodic GFN1-xTB calculations were performed using DFTB+ version 25.1 with OpenMP parallelization. Conventional SCC-DFTB calculations used the 3ob version 3.1.0, matsci version 0.3.0, and pbc version 0.3.0 Slater--Koster parameter sets \cite{elstner1998self}. Carbon was described by the corresponding $2s/2p$ atomic-orbital basis, including the overlap matrix. Periodic GFN1-xTB was evaluated through the DFTB+ tblite interface rather than the standalone xTB executable \cite{vicent2021efficient}. None of these methods was refitted to the graphene structures or to the DFT reference used in this work.}

\rev{For the energy and force benchmark, all methods were evaluated on the same 98 configurations in the graphene test split. SCC cycles were converged to a tolerance of $10^{-6}$ with at most 200 iterations, and Fermi filling at 300~K was used. Each structure used an $n_k\times n_k\times 1$ mesh, where $n_k=\max[1,\operatorname{round}(30~\text{\AA}/a)]$ and $a$ is the length of the first in-plane lattice vector. Forces were obtained analytically from the same converged calculations. For 3ob, the third-order correction was enabled with a carbon Hubbard derivative of $-0.1492$. A D3 correction with $s_6=1.0$, $s_8=0.5883$, $a_1=0.5719$, and $a_2=3.6017$ was used in the conventional DFTB calculations. Applying the same damping parameters to matsci and pbc is an explicit modeling choice in this comparison. No additional D3 correction was added to GFN1-xTB.}

\subsection{GRACE benchmark}\label{sec:method:grace}
\rev{To provide a direct comparison with an established machine-learning interatomic potential, we trained GRACE models to the same graphene energy and force data used for UnIPH. The common data split contained 1,764 training structures and 98 validation structures. The validation structures were used for checkpoint selection and are therefore data used during fitting rather than an independent test set. As summarized in Table~\ref{tab:method:grace_settings}, the comparison was approximately matched in model size and geometric resolution: both models used one configured equivariant layer, a 5.0~\angstrom\ cutoff, $\ell_{\max}=3$, float64 precision, and a maximum of 200 epochs. The GRACE model contained 94,761 trainable parameters, 6.06\% more than the verified 89,343 parameters of the graphene UnIPH model.}

\rev{\sloppy The GRACE calculations used the \texttt{GRACE\_1LAYER\_latest} preset with \texttt{n\_rad\_base} = 8, \texttt{n\_rad\_max} = 32, \texttt{prod\_func\_n\_max} = 30, \texttt{max\_order} = 4, and \texttt{n\_mlp\_dens} = 10. Model parameters were optimized with Adam using an initial learning rate of 0.008, an exponential moving average with momentum 0.99, gradient clipping at 1.0, and a cosine-decay schedule with a minimum learning rate of $6.67\times10^{-4}$. The training and validation batch sizes were 8 and 32, respectively, and random seed 0 was used. Energy and force targets used Huber losses with $\delta=0.01$, and the energy/force weights were varied as described below. Energy shifting and scaling were enabled.\par}
\rev{To assess the sensitivity to the GRACE loss weighting, we repeated the fit using six energy/force weight ratios while keeping the architecture, data split, random seed, optimizer, and 200-epoch schedule fixed. Each row in Supplementary Table~S7 was reevaluated with the checkpoint selected by the minimum validation total loss for that run. The 4:1 model was selected as the representative model because it gives the smallest maximum relative regret between the energy and force objectives.}
\rev{GRACE training was performed on an NVIDIA H100 GPU. Because the GRACE and UnIPH fits used different software implementations and training settings, their training wall times are not compared directly.}

\subsection{DeepH-E3 benchmark}\label{sec:method:deeph}
\rev{DeepH-E3 is an E(3)-equivariant neural-network framework that predicts orbital-resolved DFT Hamiltonian matrices from atomic structures \cite{deephe3}. For the present controlled comparison, we trained the DeepH-E3 architecture from scratch on the same carbon-centered scalar $p_z$-like Wannier-Hamiltonian target used for the bilayer-graphene UnIPH model. This is therefore not an evaluation of a publicly distributed pretrained DeepH-E3 model in a different atomic-orbital basis.}

\rev{Both models used the same 1,764/98/98 train/validation/test membership, a 5.0~\angstrom\ graph cutoff, one message-passing layer, and $\ell_{\max}=3$. Their capacities were also closely matched: UnIPH and DeepH-E3 contained 89,343 and 89,793 trainable parameters, respectively. The DeepH-E3 model was trained for 100 epochs, and the epoch-85 checkpoint was selected using the validation set. The test set was not used for model selection.}

\rev{Because both models learn the same orthonormal scalar $p_z$ Wannier Hamiltonian, their errors can be compared directly. We therefore evaluated the MAE and RMSE of the predicted Hamiltonian elements against the same reference for the fixed 98-structure test set.}

\subsection{Computational-time benchmarks}\label{sec:method:timing}
\rev{In panel (a), the archived UnIPH timing uses an 8,428-atom structure on an NVIDIA H100 GPU, the GRACE timing uses a 10,444-atom structure on the same NVIDIA H100 GPU, and the DFTB/matsci and GFN1-xTB timings use a 5,548-atom structure on 64 AMD EPYC 7763 CPU cores. In panel (b), UnIPH, the architecture-matched DeepH-E3 model, and LETB are evaluated for the same 10,444-atom TBG structure, while the DFTB/matsci and GFN1-xTB bars use their measured fixed-charge single-$k$ workflows. The electronic-structure timings include Hamiltonian construction and eigenvalue calculation; the DFTB/xTB calculations additionally construct the overlap matrix and solve the corresponding generalized eigenvalue problem.}

\rev{For panel (b), UnIPH, the matched DeepH-E3 model, and LETB were rerun for the same 10,444-atom structure on one NVIDIA H100 node. The two neural models used the same GPU and float64 precision, while LETB used its native implementation on the node's 64 allocated CPU cores.}

\subsection{Continuum-LJ mechanical substrate model for TBG}\label{sec:method:tbg_lj}
To account for the pulling effect exerted by the substrate, we employ a continuum Lennard-Jones (LJ) potential to model the interaction between carbon atoms and the underlying substrate. Modeling the substrate at the atomic level would require a significantly larger supercell due to the lattice structure of the substrate. To avoid this computational cost, we adopt a continuum approximation. 
\rev{The continuum LJ term is used as a computationally tractable approximation that avoids constructing a large atomistic substrate supercell and isolates the prescribed mechanical pulling component. It is an external interaction, not part of the learned UnIPH model, and contributes only to relaxation energy and forces. Because it has no corresponding degrees of freedom in the predicted Hamiltonian, the calculation retains only the geometry-mediated response of graphene and excludes charge transfer, hybridization, dielectric screening, interface dipoles, and atomistic substrate registry.} The total interaction energy $U_{\rm substrate}$ and the force $f_{\rm substrate}$ acting in the $z$-direction are given by:

\begin{equation}
U_{\rm LJ}(\rho_\mathrm{sub}, \sigma, \epsilon, z_0) =
\frac{4}{3} \pi \rho_\mathrm{sub} \epsilon \sigma^3
\left[
\frac{1}{15} \left( \frac{\sigma}{z_0} \right)^9
- \frac{1}{2} \left( \frac{\sigma}{z_0} \right)^3
\right],
\end{equation}
and
\begin{equation}
f_{\rm LJ}(\rho_\mathrm{sub}, \sigma, \epsilon, z_0) =
\frac{4}{3} \pi \rho_\mathrm{sub} \epsilon \sigma^2
\left[
\frac{3}{5} \left( \frac{\sigma}{z_0} \right)^{10}
- \frac{3}{2} \left( \frac{\sigma}{z_0} \right)^4
\right],
\end{equation}

where $\rho_\mathrm{sub}$ denotes the atomic number density of the substrate, $\epsilon$ and $\sigma$ are the LJ parameters characterizing the strength and range of the interaction, respectively, and $z_0$ is the vertical distance between a carbon atom and the substrate surface.
In our simulations, we postulated a continuum Si$_3$N$_4$ substrate where only the C–Si and C–N interactions were considered \cite{choi2024}. \rev{We selected it as the reference because it is commonly used in TBG experiments and its reported LJ parameters produce a stronger mechanical pulling effect than hBN \cite{choi2024}.} The corresponding LJ parameters are listed in \tab{tab:method:tbg_lj}.
The atomic number densities used are $\rho_{\rm Si} = 0.042~\angstrom^{-3}$ and $\rho_{\rm N} = 0.056~\angstrom^{-3}$.
External energy and force applied to each carbon atom during structural relaxation ($U_{\rm substrate}(z_0)$ and $f_{\rm substrate}(z_0)$) are given by:
\begin{equation}
    U_{\rm substrate}(z_0) = U_{\rm LJ}(\rho_\mathrm{Si}, \sigma_{\rm C-Si}, \epsilon_{\rm C-Si}, z_0) + U_{\rm LJ}(\rho_\mathrm{N}, \sigma_{\rm C-N}, \epsilon_{\rm C-N}, z_0),
\end{equation}
\begin{equation}
    f_{\rm substrate}(z_0) = f_{\rm LJ}(\rho_\mathrm{Si}, \sigma_{\rm C-Si}, \epsilon_{\rm C-Si}, z_0) + f_{\rm LJ}(\rho_\mathrm{N}, \sigma_{\rm C-N}, \epsilon_{\rm C-N}, z_0).
\end{equation}

\subsection{Continuum-LJ substrate and local electronic mapping for monolayer MoS\texorpdfstring{\textsubscript{2}}{2}}\label{sec:method:mos2_substrate}
The substrate-induced deformation calculations in \rev{\fig{fig:MOS2:LDOS}} were performed for a $20\times20\times1$ monolayer MoS$_2$ supercell containing 1200 atoms. The relaxed monolayer was first shifted vertically so that the bottom S layer was initially placed 5.0\,\angstrom\ above the substrate reference plane, and the structure was then relaxed on a periodic continuum substrate using the interatomic potential predicted by UnIPH together with a substrate interaction term. Following the tangent-plane treatment used in atomistically informed continuum modeling of 2D materials on patterned substrates \cite{choi2023atomistically}, the substrate geometry was evaluated in a periodic form using minimum-image in-plane distances.
\rev{The LJ term is applied outside UnIPH and supplies only mechanical energy and forces; the continuum normal-force approximation does not resolve in-plane atomistic registry.}

The substrate height field was defined on the same in-plane periodic cell as the MoS$_2$ supercell. For an atom at in-plane position $\mathbf{r}_i=(x_i,y_i)$ and feature center $\mathbf{r}_c$, the minimum-image radial displacement from the feature center was $\Delta \mathbf{r}_i=\Delta \mathbf{r}_{\min}(\mathbf{r}_i-\mathbf{r}_c)$ and $r_i=|\Delta \mathbf{r}_i|$. The substrate profile was then written as
\begin{equation}
z_{\mathrm{sub}}(r)=
\begin{cases}
\dfrac{A}{4}\left[\cos\left(\dfrac{\pi r}{R}\right)+1\right]^2, & r<R,\\[6pt]
0, & r \ge R,
\end{cases}
\end{equation}
where $A$ is the substrate amplitude and $R$ is the feature radius. The feature center was placed at fractional coordinates $(0.5,0.5)$ and the radius was fixed to $R=0.10|\mathbf{a}_1|=6.36$\,\angstrom. Three substrate amplitudes were examined: $A=0$, 1, and 2\,\angstrom.
The in-plane gradient was evaluated as
\begin{equation}
\nabla z_{\mathrm{sub}}(\mathbf{r}_i)=
\left.\frac{dz_{\mathrm{sub}}}{dr}\right|_{r_i}
\frac{\Delta \mathbf{r}_i}{r_i},
\end{equation}
with $\nabla z_{\mathrm{sub}}=\mathbf{0}$ at $r_i=0$ and outside the feature. Writing $\nabla z_{\mathrm{sub}}=(z_x,z_y)$, the local substrate normal vector was
\begin{equation}
\mathbf{n}_i=\frac{1}{\sqrt{1+z_x^2+z_y^2}}
\left(-z_x,-z_y,1\right).
\end{equation}
The substrate surface point directly below atom $i$ was $\mathbf{x}_{\mathrm{sub},i}=(x_i,y_i,z_{\mathrm{sub}}(x_i,y_i))$, and the distance entering the continuum interaction was taken as the tangent-plane normal projection
\begin{equation}
h_i=\left(\mathbf{x}_i-\mathbf{x}_{\mathrm{sub},i}\right)\cdot\mathbf{n}_i,
\end{equation}
which reduces to the usual vertical separation for a flat substrate.

The interaction between each MoS$_2$ atom and the substrate was described by the half-space Lennard-Jones approximation \cite{choi2023atomistically}
\begin{equation}
\psi_{\rm{X-B}}(h)=\frac{6}{5}\pi \epsilon_{\rm{X-B}}n_{\rm B}\sigma_{\rm{X-B}}^2
\left(
\frac{2\sigma_{\rm{X-B}}^{10}}{27h^9}
-
\frac{5\sigma_{\rm{X-B}}^4}{9h^3}
\right),
\end{equation}
where $h$ is the local tangent-plane distance, $\epsilon_{\rm X-B}$ and $\sigma_{\rm X-B}$ are the Lennard-Jones parameters for atom species $X$ in MoS$_2$ and substrate species $B$, and $n_{\rm B}$ is the substrate number density. The LJ parameters used in the present calculations are summarized in \tab{tab:method:mos2_lj}.

The substrate contribution for each atom was obtained by summing the Si and N terms,
\begin{equation}
U_i^{\mathrm{sub}}=\psi_{\rm{X_i}-\mathrm{Si}}(h_i)+\psi_{\rm{X_i}-\mathrm{N}}(h_i),
\end{equation}
and the corresponding substrate force was applied along the local surface normal as
\begin{equation}
\mathbf{F}_i^{\mathrm{sub}}=
-\left.\frac{dU_i^{\mathrm{sub}}}{dh}\right|_{h=h_i}\mathbf{n}_i,
\end{equation}
where $X_i$ is the MoS$_2$ atom species and $\mathbf{n}_i$, $h_i$, and $\left.dU_i^{\mathrm{sub}}/dh\right|_{h=h_i}$ were updated at every calculator evaluation using the current relaxed geometry. \rev{This normal-force approximation retains the prescribed mechanical pulling component of the smooth substrate while accounting for the local surface orientation \cite{choi2023atomistically}.} Structural relaxation was carried out with the FIRE algorithm until the maximum residual force fell below $10^{-6}$\,eV/\angstrom, using a maximum of 5000 steps.

After relaxation, the strain field was evaluated using only the Mo sublattice, which contains 400 atoms in the $20\times20$ supercell. For each Mo atom, a local affine deformation gradient $\mathbf{F}^{(a)}$ was obtained from the six nearest Mo neighbors by comparing the relaxed geometry with the pre-substrate reference supercell. The in-plane strain tensor was then computed as
\begin{equation}
\boldsymbol{\varepsilon}^{(a)}=\frac{1}{2}\left(\mathbf{F}^{(a)}+\mathbf{F}^{(a)\mathsf{T}}\right)-\mathbf{I},
\end{equation}
and the largest principal value of $\boldsymbol{\varepsilon}^{(a)}$ was taken as the local maximum principal strain. To visualize the spatial distribution, the atom-resolved strain values were mapped onto a periodic $160\times160$ real-space grid using Gaussian smoothing with $\sigma=2.5$\,\angstrom.

\subsubsection{LDOS, local band edges, and band-gap mapping}
\rev{The electronic calculation contains only the Mo- and S-centered degrees of freedom represented by the trained UnIPH Hamiltonian. The continuum substrate affects the result only through the relaxed coordinates; no substrate-induced charge redistribution, hybridization, or dielectric renormalization is included. The LDOS and band-edge maps are therefore conditional mechanical-proxy predictions.}
The local electronic structure was evaluated from the UnIPH-predicted tight-binding Hamiltonian of the relaxed supercell. For each atomic configuration, the model generated local tight-binding blocks, which were assembled into a Bloch Hamiltonian
\begin{equation}
H(\mathbf{k})=\sum_{\mathbf{R}} H(\mathbf{R}) e^{i 2\pi \mathbf{k}\cdot \mathbf{R}},
\end{equation}
and diagonalized on a $5\times5\times1$ Brillouin-zone mesh. The local density of states (LDOS) was evaluated on an energy grid from $-3.5$ to $-0.5$\,eV using 2000 energy points and Gaussian broadening $\eta=0.1$\,eV.
For atom $a$, the atom-resolved LDOS was defined as
\begin{equation}
\rho_a(E)=\frac{1}{N_k\eta\sqrt{2\pi}}\sum_{\mathbf{k},n} w_{a,n\mathbf{k}}
\exp\left[-\frac{(E-\varepsilon_{n\mathbf{k}})^2}{2\eta^2}\right],
\end{equation}
where $w_{a,n\mathbf{k}}$ is the orbital weight of eigenstate $n\mathbf{k}$ on atom $a$. The atom-resolved LDOS values from all atoms were then projected onto a periodic $80\times80$ real-space grid by Gaussian spatial averaging. For a grid point $\mathbf{r}$, the spatial weights were
\begin{equation}
W_a(\mathbf{r})=\exp\left[-\frac{d_a^2(\mathbf{r})}{2\sigma_r^2}\right],\qquad
\widetilde{W}_a(\mathbf{r})=\frac{W_a(\mathbf{r})}{\sum_b W_b(\mathbf{r})},
\end{equation}
with minimum-image in-plane distance $d_a(\mathbf{r})$ and smoothing width $\sigma_r=4.0$\,\angstrom. The resulting grid LDOS field was
\begin{equation}
\rho(\mathbf{r},E)=\sum_a \widetilde{W}_a(\mathbf{r})\rho_a(E).
\end{equation}
The local band edges were extracted directly from the grid LDOS. At each real-space grid point, energies satisfying
\begin{equation}
\rho(\mathbf{r},E)\le \alpha \rho_{\max}(\mathbf{r}),
\end{equation}
with $\alpha=0.01$, were identified as a low-LDOS window. Among all such windows, the widest interval was taken as the local gap region; its lower edge was assigned to the local valence band maximum (VBM), its upper edge to the local conduction band minimum (CBM), and their difference to the local band gap. The resulting local VBM, local CBM, and local band-gap fields were written on the $80\times80$ grid. For the line profiles in \rev{\fig{fig:MOS2:LDOS}b}, the local VBM and CBM were sampled along the supercell diagonal from $(u,v)=(0,1)$ to $(1,0)$ in fractional coordinates using 80 equally spaced points, corresponding to a total line length of 110.22\,\angstrom.

\subsection{\rev{Use of generative AI}}
\rev{During manuscript revision, the authors used OpenAI Codex (GPT-5) to assist with language editing and code editing. All outputs were reviewed and verified by the authors, who take full responsibility for the final content.}

\section*{Data Availability}
The datasets generated and/or analysed during the current study are available at \href{https://zenodo.org/records/20500533}{https://zenodo.org/records/20500533}. 
\section*{Code Availability}
The underlying code for this study can be accessed via this link \href{https://github.com/Moon-ki8/UnIPH}{https://github.com/Moon-ki8/UnIPH}.

\section*{Acknowledgements}
This work used Delta at National Center for Supercomputing Applications through allocation MAT240019 from the Advanced Cyberinfrastructure Coordination Ecosystem: Services \& Support (ACCESS) program, which is supported by National Science Foundation grants \#2138259, \#2138286, \#2138307, \#2137603, and \#2138296.

\section*{Funding}
This work was supported by the Air Force Office of Scientific Research (AFOSR), USA through the Multidisciplinary University Research Initiative (MURI) Award number FA9550-23-1-0334. The funder had no role in study design, data collection and analysis, decision to publish, or preparation of the manuscript.

\section*{\rev{Author Contributions}}
\rev{M.-K.C. performed the simulations, collected and analyzed the data, developed the machine-learning model, applied the machine-learning model to 2D materials, and wrote the manuscript. D.P. contributed to dataset construction and reviewed the manuscript. S.S. reviewed the code and the manuscript. H.T.J. supervised the project, acquired funding, and reviewed and revised the manuscript.}

\section*{Competing Interests}
\rev{The authors declare no competing financial or non-financial interests.}

\bibliography{main}

\clearpage
\section*{Figure legends}

\begin{figure}[h!]
\includegraphics[width=0.95\textwidth]{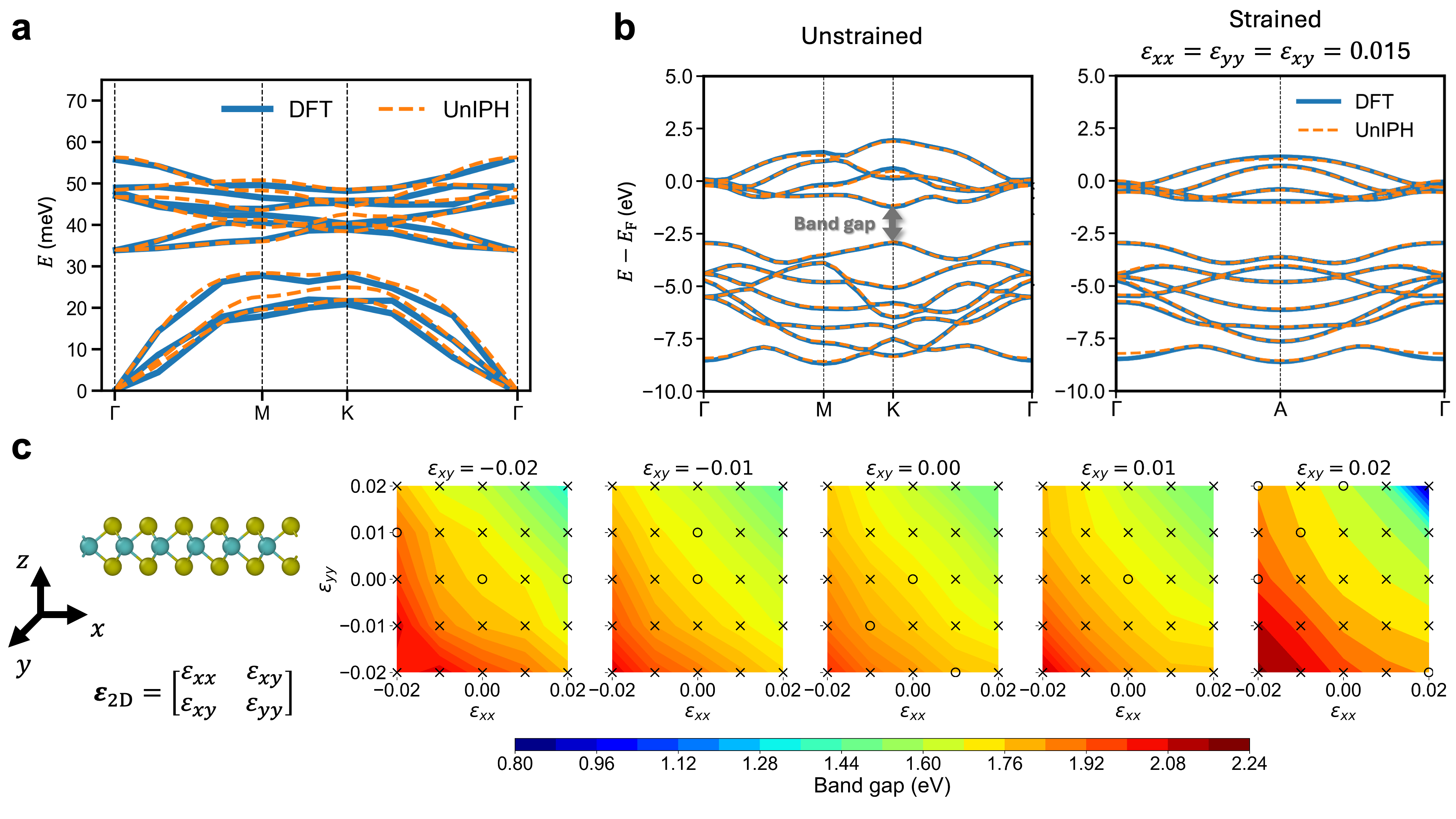}
\caption{Validation of UnIPH for IP and TB Hamiltonians of monolayer MoS$_2$.
(a) Phonon dispersion relations from UnIPH and DFT along the high-symmetry path $\Gamma$--M--K--$\Gamma$.
(b) Electronic band structures from UnIPH and DFT for unstrained and uniformly strained MoS$_2$ ($\varepsilon_{xx} = \varepsilon_{yy} = \varepsilon_{xy} = 0.015$), with energies referenced to the Fermi level ($E_\mathrm{F} = 0$).
(c) Band gap evolution under uniform 2D strain in $\varepsilon_{xx}$ and $\varepsilon_{yy}$ for various fixed values of $\varepsilon_{xy}$. Open circles and crosses indicate direct and indirect band gaps, respectively. \rev{Molecular structures in this figure were visualized using OVITO version 3.7.6.}}
\label{fig:MOS2:IPTB}
\end{figure}

\begin{figure}[h!]
\includegraphics[width=1.0\textwidth]{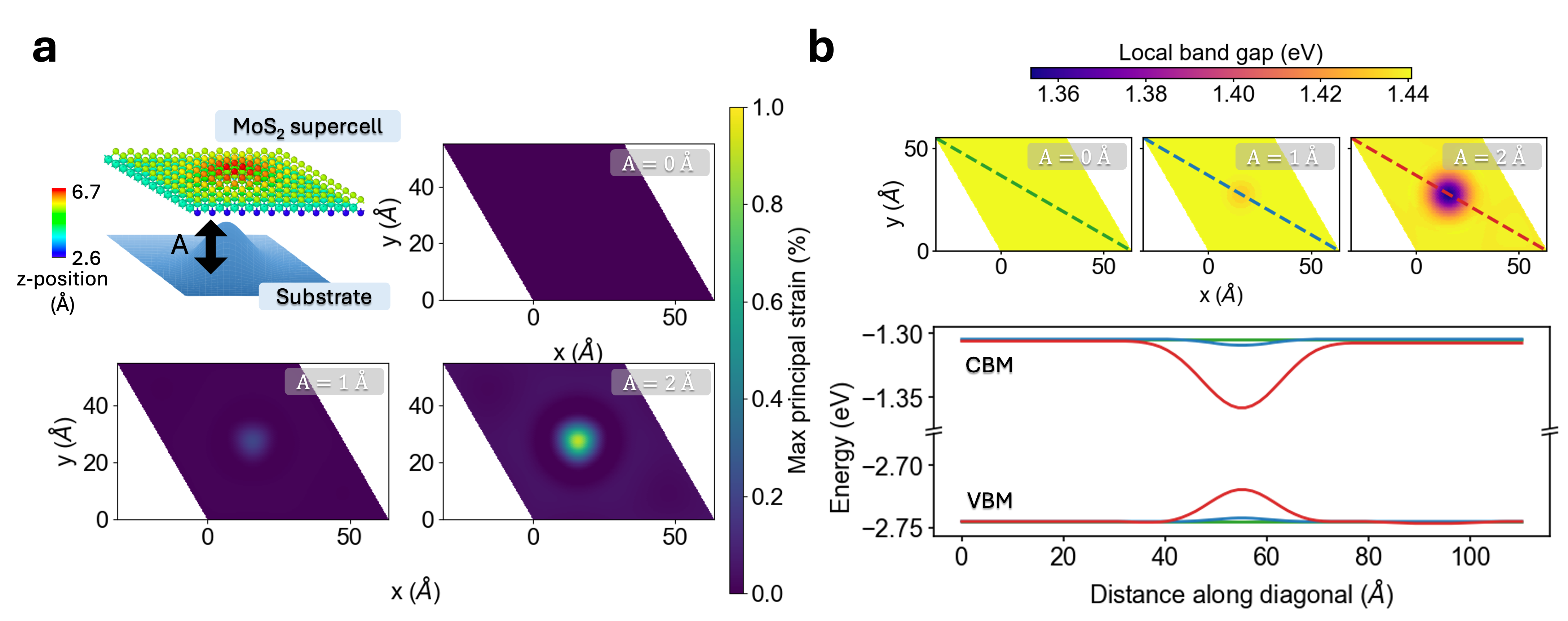}
\caption{\rev{Strain-mediated band-edge modulation in monolayer MoS$_2$ under a continuum-LJ mechanical substrate model.}
(a) Relaxed $20\times20$ monolayer MoS$_2$ supercells on a non-uniform substrate, together with the corresponding maximum principal strain maps for $A=0$, 1, and 2\,\angstrom.
(b) Local band-gap maps for the three substrate amplitudes ($A=0$, 1, and 2\,\angstrom), followed below by diagonal line cuts of the local conduction band minimum (CBM) and valence band maximum (VBM). \rev{Molecular structures in this figure were visualized using OVITO version 3.7.6.}}
\label{fig:MOS2:LDOS}
\end{figure}

\clearpage
\section*{Tables}

\begin{table}[ht!]
\centering
\footnotesize
\begin{tabular}{@{}lccc@{}}
\toprule
\textbf{System} & \textbf{Cutoff Distance (\angstrom)} & $\bm{\ell_{\max}}$ & \textbf{Irreps} \\
\midrule
\multirow{2}{*}{Bilayer Graphene} 
& \multirow{2}{*}{5.0} 
& \multirow{2}{*}{3} 
& \textbf{Node:} \texttt{14x0o+14x0e+14x1o+14x1e} \\
& & & \texttt{+6x2o+6x2e+6x3o+6x3e} \\
& & & \textbf{Edge:} \texttt{5x0o+5x0e+2x1o+2x1e+1x2o+1x2e+1x3o+1x3e} \\
\midrule
Monolayer MoS$_2$ 
& 7.0 
& 4 
& \textbf{Node \& Edge:} \texttt{15x0e+15x0o+8x1e+8x1o} \\
& & & \texttt{+8x2e+8x2o+8x3e+8x3o} \\
\bottomrule
\end{tabular}
\caption{Graph construction parameters and irreducible representations (irreps) for each material system.}
\label{tab:method:gnn_parameter}
\end{table}

\begin{table}[h!]
\centering

\begin{tabular}{lccc}
\hline
Material & $\langle E \rangle$ (eV) & $\sigma_f$ (eV/\AA) & $\sigma_{\mathrm{TB}}$ (eV) \\
\hline
Bilayer graphene & $-250.984642$ & $0.842210$ & $0.265346$ \\
Monolayer MoS$_2$ & $-2252.602666$ & $0.607018$ & $1.000000$ \\
\hline
\end{tabular}
\caption{
Normalization constants for bilayer graphene and monolayer MoS$_2$. 
$\langle E \rangle$ is the mean per-atom energy, $\sigma_f$ is the RMS magnitude of atomic forces, and $\sigma_{\mathrm{TB}}$ is the RMS of TB Hamiltonian elements. For monolayer MoS$_2$, $\sigma_{\mathrm{TB}}$ is fixed to $1.0$ eV as a predefined scaling factor.
}
\label{tab:method:normalization_values}
\end{table}

\begin{table}[!ht]
\centering
\begin{tabular}{@{}llp{6.6cm}r@{}} 
\hline
System & Supercell & Condition & \# Samples \\
\hline\hline
\multirow{7}{*}{Bilayer graphene}
 & 1$\times$1 & Interlayer shift; $d_z$ variation; in-plane strain ($\le$ 2\%) & 550 \\
 & 3$\times$3 & Random displacement ($\le$ 0.03\,\angstrom) & 205 \\
 & 3$\times$3 & AIMD at T=50~K & 100 \\
 & 4$\times$4 & AIMD at T=200~K & 141 \\
 & 1$\times$1 & Interlayer shift; $d_z$ variation; no strain & 540 \\
 & 1$\times$1 & Interlayer shift; $d_z$ variation; in-plane strain ($\le$ 3\%) & 300 \\
 & 3$\times$3 & Random displacement ($\le$ 0.03\,\angstrom) with grid sampling & 124 \\
\hline
\multicolumn{3}{r}{\textbf{Total}} & \textbf{1960} \\
\hline
\multirow{5}{*}{Monolayer MoS$_2$}  & 1$\times$1 & in-plane strain ($\le$  2\%); \newline thickness variation ($\le$ 1\%); \newline random displacement ($\le$  0.02\,\angstrom) & 200 \\
  & 4$\times$4 & AIMD at 50~K & 100 \\
  & 4$\times$4 & AIMD at 100~K & 100 \\
  & 4$\times$4 & AIMD at 200~K & 100 \\
  & 4$\times$4 & AIMD at 300~K & 100 \\
\hline
\multicolumn{3}{r}{\textbf{Total}} & \textbf{600} \\
\hline
\end{tabular}
\caption{Details of configurations for bilayer graphene and monolayer MoS$_2$ for DFT+Wannier90 calculation.}
\label{tab:method:dft_dataset}
\end{table}

\begin{table}[h!]
\centering
\small
\begin{tabular}{lcc}
\toprule
Setting & UnIPH & GRACE \\
\midrule
Configured equivariant layers & 1 & 1 \\
Cutoff (\angstrom) & 5.0 & 5.0 \\
$\ell_{\max}$ & 3 & 3 \\
Trainable parameters & 89,343 & 94,761 \\
Training/validation structures & 1,764/98 & 1,764/98 \\
Maximum epochs & 200 & 200 \\
Precision & float64 & float64 \\
\bottomrule
\end{tabular}
\caption{Conditions used for the approximately matched UnIPH--GRACE graphene comparison. The implementations retain their native model definitions, optimizers, and loss functions.}
\label{tab:method:grace_settings}
\end{table}

\begin{table}[ht!]
\centering
\begin{tabular}{ccc}
\hline
Interaction & \(\epsilon\) (meV) & \(\sigma\) (nm) \\
\hline
C--Si & 10.46 & 0.39 \\
C--N & 21.36 & 0.37 \\
\hline
\end{tabular}
\caption{LJ parameters for the interaction of carbon with substrate atoms Si and N.}\label{tab:method:tbg_lj}
\end{table}

\begin{table}[ht!]
\centering
\begin{tabular}{lccc}
\toprule
Interaction & $\epsilon$ (eV) & $\sigma$ (\angstrom) & $n_B$ (\angstrom$^{-3}$) \\
\midrule
Mo--Si & 0.038120 & 3.3022 & 0.0418552 \\
Mo--N  & 0.077892 & 3.0261 & 0.0558070 \\
S--Si  & 0.001799 & 3.7106 & 0.0418552 \\
S--N   & 0.003672 & 3.4345 & 0.0558070 \\
\bottomrule
\end{tabular}

\caption{Lennard-Jones parameters used for the MoS$_2$--substrate interaction in \rev{\fig{fig:MOS2:LDOS}}.}
\label{tab:method:mos2_lj}
\end{table}

\end{document}